\documentclass[apj]{emulateapj}
\usepackage{graphicx,hyperref}
\usepackage{txfonts}
\usepackage{epstopdf}
\usepackage{longtable}
\usepackage{natbib}
\usepackage{color}
\newcommand{\scs}{\scriptsize}

\shorttitle{Solar twins and the Barium puzzle}
\shortauthors{Reddy and Lambert}

\begin{document} 

\title{Solar twins and the Barium puzzle}
\author{Arumalla B. S. Reddy}
\author{David L. Lambert}
\affil{W.J. McDonald Observatory and Department of Astronomy, The University of Texas at Austin, Austin, TX 78712-1205, USA}
\email{bala@astro.as.utexas.edu}

\voffset=-18.5mm


\begin{abstract}

Several abundance analyses of Galactic open clusters (OCs) have shown a tendency for Ba but not for other heavy elements (La$-$Sm) to increase sharply with decreasing age such that Ba was claimed to reach [Ba/Fe] $\simeq +0.6$ in the youngest clusters (ages $<$ 100 Myr) rising from [Ba/Fe]$=0.00$ dex in solar-age clusters. Within the formulation of the $s$-process, the difficulty to replicate higher Ba abundance and normal La$-$Sm abundances in young clusters is known as {\it the barium puzzle}. Here, we investigate the barium puzzle using extremely high-resolution and high signal-to-noise spectra of 24 solar twins and measured the heavy elements Ba, La, Ce, Nd and Sm with a precision of 0.03 dex.  We demonstrate that the enhanced Ba {\scs II} relative to La$-$Sm seen among solar twins, stellar associations and OCs at young ages ($<$100 Myr) is unrelated to aspects of stellar nucleosynthesis but has resulted from overestimation of Ba by standard methods of LTE abundance analysis in which the microturbulence derived from the Fe lines formed deep in the photosphere is insufficient to represent the true line broadening imposed on Ba {\scs II} lines by the upper photospheric layers from where the Ba {\scs II} lines emerge. As the young stars have relatively active photospheres, Ba overabundances most likely result from the adoption of too low a value of microturbulence in the spectum synthesis of the strong Ba {\scs II} lines but the change of microturbulence in the upper photosphere has only a minor affect on La$-$Sm abundances measured from the weak lines.

\end{abstract}


\section{Introduction} 

\begin{figure}
\begin{center}
\includegraphics[trim=0.1cm 9.9cm 6.7cm 4.6cm, clip=true,height=0.25\textheight,width=0.7\textwidth]{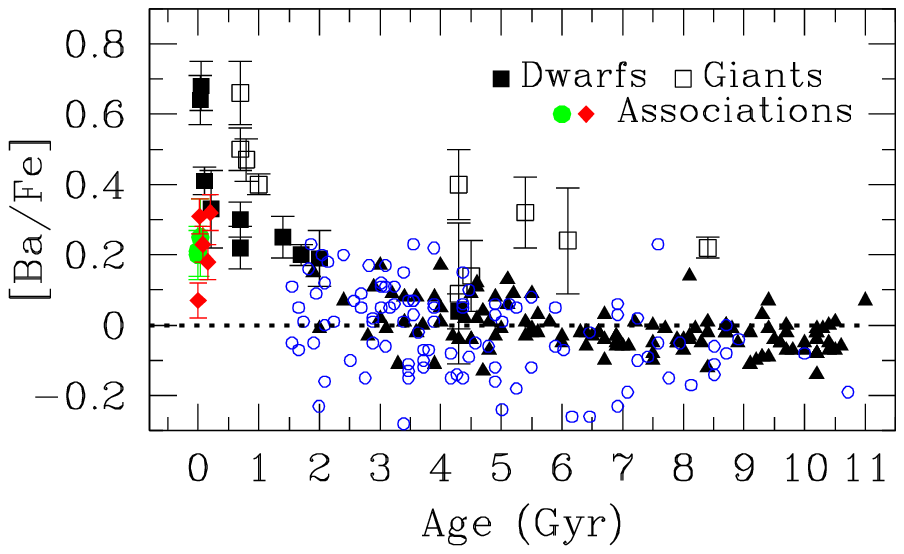}
\caption[]{The average [Ba/Fe] versus age for dwarfs and giants in OCs \citep{dorazi09} and for the Galactic disk field dwarfs from \cite[][open circles]{edvardsson93} and \cite[][filled triangles]{bensby14} covering the [Fe/H] range $-$0.3 to $+$0.3 dex of the OCs. The average [Ba/Fe] values for the stellar associations analysed by \cite{dorazi12} and \cite{redlam15} are shown as filled circles and filled rhombuses, respectively.  }
\label{baage}
\end{center}
\end{figure}

Chemical Evolution of the Galaxy (GCE) is a topic rich in puzzles that demands observers must continually provide new data to constrain models of stellar nucleosynthesis and stellar evolution. One such contemporary puzzle is the surprising claim by \cite{dorazi09} that the barium abundance in open clusters (OCs) increases with decreasing age of the cluster. 
The standard local thermodynamic equilibrium (LTE) abundance analysis of the Ba\,{\sc ii} 5853 \AA\ and 6496 \AA\ lines in FG dwarfs in 10 OCs with ages from 35 Myr to 8.4 Gyr and red giants in 10 OCs older than 700 Myr gave [Ba/Fe] $\sim$ $+$0.6 dex in the youngest clusters (age $<$ 100 Myr) decreasing to $0.0$ dex in clusters older than the Sun. The [Ba/Fe] values are systematically higher in clusters analysed for giants (see Figure \ref{baage}). 
This apparent overabundance of [Ba/Fe] in young clusters is referred to as {\it the barium puzzle}. 
Figure \ref{baage} constructed from the [Ba/Fe] values in OCs \cite{dorazi09} and FG field dwarfs \citep{edvardsson93,bensby14} illustrates the barium puzzle. 

Another dimension to the puzzle was provided by \cite{dorazi12} who analysed F and G dwarfs in three associations (ages 50-500 Myr) finding that Y, Zr, La and Ce relative to Fe had solar ratios except for Ba which is overabundant but by only about 0.2 dex. Our abundance analysis of FGK dwarfs \citep{redlam15} in five local associations (Orion, Argus, Subgroup B4, Hercules-Lyra and Carina-Near) spanning the age range 3 Myr for Orion \citep{walker1969,blaauw1954} to 200 Myr for Carina-Near \citep{zuckerman06} supports \cite{dorazi12}'s results and confirms solar abundance ratios for all elements (Na$-$Ni, Y, Zr, La$-$Eu) but for Ba which was overabundant by $+0.07$ dex (Orion) to $+0.32$ dex (Argus, Carina-Near). The fact that the Ba abundance for the youngest solar metallicity associations is much less than the value $\sim\,$0.6 dex reported for the OCs IC 2391 and IC 2602 of ages $<$ 50 Myr is another aspect of the barium puzzle \citep[Figure \ref{baage}, see also the Figure 6 in][]{dorazi12}. How can associations and OCs of comparable age have very different chemical compositions\,?

Within the constraints of the $s$-process operation in AGB stars, the higher Ba yield should be accompanied by higher yields of other $s$-process elements La and Ce, also Nd, and Sm \citep{busso99}. However to account for the high Ba abundance in OCs, \cite{mishenina15} reintroduced the intermediate neutron-capture $i$-process discussed earlier by \cite{cowan1977} in which the neutron density operating the process is approximately 10$^{15}$ cm$^{-3}$. i.e., intermediate between the low densities for the $s$-process and the much higher densities necessary for the $r$-process. But the abundance predictions from the $i$-process (Figure 6 in \cite{mishenina15}) are very much different from those measured in OCs and associations: [Ba/La] $\simeq 1.2$ with [Eu/La] $\simeq -0.5$ are predicted but the results for OCs and associations are [Ba/La] $\sim 0.4$ and $\sim 0.2$, respectively with [Eu/La] near $0.00$. 

The difficulty in replicating [Ba/La] values of about $+0.2$ to $+0.4$ dex by operation of the $s$-process in AGB stars \citep{busso99} and the large increase in Ba abundance in a {\it very} short Galactic time scale ($<$\,100 Myr) encourages one to solve the Ba puzzle not through stellar nucleosynthesis but through alterations to the standard model atmospheres and line formation including the simple device of adapting the microturbulence \citep[see][]{redlam15}. 

Our attempt \citep{redlam15} to resolve the Ba overabundance in young OCs and local associations using the Ba\,{\sc I} 5535 \AA\ line was limited by the strong temperature sensitivity of the Ba {\scs I} line that returned a solar Ba abundance for stars with effective temperature of the Sun but increasingly subsolar Ba abundances for cooler stars with apparent Ba deficiencies of 0.5 dex at 5100 K. This disturbing trend with temperature was interpreted as a strong non-LTE effect on the Ba\,{\sc i} line. Here, a reassesment of the Ba {\scs I} line's contribution to the spectrum of FGK dwarfs concludes that the contribution is minimal and, thus, contrary to \cite{redlam15} the Ba {\scs I} line is not a useful abundance indicator. 

In this paper, a sample of solar twins is analysed and provides an excellent opportunity to overcome all sorts of systematic uncertainties including the non-LTE effects. One such sample with ratios of [Ba/Fe] correlating well with stellar age was reported recently by \cite{nissen16} for 21 solar twins.
In addition, we included three solar twins (HD 42807, HD 59967 and HD 202628) whose 
stellar parameters and [Fe/H] values determined using the 1D-LTE {\scs \bf MARCS} models are available from \cite{ramirez14}.

Twenty-one solar twins chosen for abundance analysis by \cite{nissen15,nissen16} were taken from the analysis of HARPS spectra \citep{mayor03} by \cite{sousa08}. The stars cover the age range 0.4 to 10 Gyr with spectra having a signal-to-noise ratio (S/N) $\geq$ 600 and the stellar parameters within $\pm$100 K in effective temperature, T$_{\rm eff}$, $\pm0.15$ dex in surface gravity, log~$g$, and $\pm$0.11 dex in metallicity, [Fe/H], about the solar values. By exploiting these extremely high S/N and resolution ($R\,\simeq$\,115,000) HARPS spectra, T$_{\rm eff}$ and log~$g$ were reestimated by \cite{nissen15} who also measured chemical abundances for elements C, O, Na, Mg, Al, Si, S, Ca, Sc, Ti, Cr, Mn, Fe, Ni, Cu, Zn, Y, and Ba by \citep{nissen15,nissen16}. The estimated 1-sigma errors in stellar parameters are $\sigma(T_{\rm eff})\sim \pm$6 K, $\sigma(log~$g$)\sim \pm$0.012 dex, and abundances were measured with a typical precision of $\pm$0.01 dex \citep{ramirez14,nissen16}. The primary stimulus for this paper was Nissen's  reported increase in [Ba/Fe] with decreasing age \citep[Figure 4 of][]{nissen16} such that Ba was claimed to reach [Ba/Fe] $\simeq +0.2$ for stars younger than 6 Gyr and flatten out at a level of $-$0.03 dex for stars older than 6 Gyr. 

In this paper, we expand the the information available on heavy elements in solar twins to the suite of elements La, Ce, Nd, Sm. This is the first time that the solar twins have been exploited for a thorough investigation of the Ba puzzle via the analysis of lines of Ba and other heavy elements. 
More importantly, we examine if the abundances Ba$-$Sm across the sample of solar twins depend on stellar activity as measured by the Ca {\scs II} H and K index log\,R$^{\prime}_{\rm HK}$. Stars with active chromospheres are likely to have an upper photosphere differing from that predicted by a classical model atmosphere and, hence, strengths of strong lines such as the Ba\,{\sc ii} lines may be incorrectly modeled. We compare the results for solar twins with measurements for FGK dwarfs in OCs and stellar associations in the Galactic disk. 

The layout of the paper is as follows: In Section 2 we describe the assembly of the spectra, LTE analysis of the Ba {\scs II} 5853 \AA\ and lithium lines and explore the [Ba/Fe] and log\,$\epsilon$(Li) versus age of the solar twins. Section 3 presents the abundance analysis of elements La, Ce, Nd, Sm and their variation with stellar age, and closes with a short discussion on the barium abundance in relation to other heavy elements. In Section 4 we investigate the relation of heavy elements in the solar twins, associations and OCs to the stellar chromospheric activity. Finally, in Section 5 we provide our conclusions concerning the major advancements of this paper in the resolution of the Ba puzzle in OCs and stellar associations.

\section{The sample of solar twins}
Multiple reduced spectra of 24 solar twins and the Sun (reflected light from the asteroid Vesta) were retrieved from the ESO public archive with program IDs 072.C-0488, 074.C-0364, 088.C-0323, 183.C-0972, 188.C-0265, 192.C-0224. All these targets were observed with the HARPS spectrograph \citep{mayor03} at the 3.6-m telescope of the La Silla observatory whose spectra corresponds to a resolution of $R\,\simeq$\,115,000. Multiple spectra of a given star were Doppler corrected and coadded to give a spectrum with S/N$>$600 for all solar twins while the S/N ratio of the solar spectrum equals about 1100. All these targets are confirmed solar twins covering a limited range of 5690 to 5871 K in T$_{\rm eff}$, 4.25 to 4.54 in log~$g$, $\pm$0.11 dex in [Fe/H] and the stellar masses similar to the Sun (M\,$=$\,1.0$\pm$0.07 M$_{\odot}$) but having a range in ages from 0.4 to 10 Gyr and log\,R$^{\prime}_{\rm HK}$ values of $-$5.0 to $-$4.4. Adopting the set of MARCS model atmospheres \citep{gustafsson08} and the Uppsala BSYN program, \cite{nissen15,nissen16} measured the chemical abundances of 21 stars for many elements including two $s$-process elements Y and Ba. Details of observations and chemical abundances of many elements will be found in those papers. The three solar twins (HD 42807, HD 59967 and HD 202628) taken from \cite{ramirez14} have stellar parameters and only [Fe/H] values determined using the 1D-LTE {\scs \bf MARCS} models. In this paper, we extend the abundance analysis to Li and heavy elements Ba, La, Ce, Nd and Sm in 24 solar twins.

\subsection{The Ba\,{\sc i} 5535 \AA\ line} 

In cool dwarfs and giants, barium is heavily ionized and the Ba\,{\sc ii} lines are normally very strong and, as a result, the Ba abundance is uncertain. An alternative determination of the Ba abundance from Ba\,{\sc i} lines would be of great value. We \citep{redlam15} used the Ba\,{\sc i} resonance line at 5535.481 \AA\ to determine the Ba abundance in FGK dwarfs following a synthesis of the line in the solar spectrum. Synthesis is demanded because the Ba\,{\sc i} line's contribution falls in the red wing of a strong Fe\,{\sc i} line; a potential Ba\,{\sc i} signature is seen as an inflection in the Fe\,{\sc i} line's red wing. In our 2015 syntheses, the $gf$-value of the Ba\,{\sc i} line was adjusted so that the solar 5535 \AA\ blend was fitted with the Ba abundance set at the solar value. This `astrophysical' gf-value when adopted for stars gave sensible results for dwarfs with effective temperatures greater than about 5800 K. The 5535 \AA\ line in cooler stars gave increasingly subsolar Ba abundances reaching $-0.5$ dex at about 5100 K. 
  
On reflection, the 2015 synthesis and its application is seriously flawed by the misidentification of the solar line as due to Ba {\scs I}. Most critically, the $gf$-value of 5535 \AA\ Ba\,{\sc i} resonance line is determined by experiment to 1-2 \% \citep{kelly1977,klose02,curry04} to be $\log gf = 0.215$ but the 2015 solar synthesis adopting the inverted solar $gf$-value required an increment of almost 1 dex over the $gf$-value estimated from the experiments. The obvious conclusion must be that another and significant line is present in the Fe\,{\sc i} -- Ba\,{\sc i} blend. A search of the usual sources has not uncovered the contributing atom, ion or molecule. Kurucz's list\footnote{\url{http://kurucz.harvard.edu/linelists.html}} offers two very weak excited Fe\,{\sc i} lines as possibilities but their $gf$-values have to be increased at least a 1000-fold to match the solar spectrum when the Ba\,{\sc i} line given its correct strength. Introduction of an Fe\,{\sc i} line improves the fit to the blend across the effective interval 5000 to 6500 K but the impression is that the unidentified line has a higher excitation potential than the lines from Kurucz. Given that the Ba\,{\sc i} resonance line makes only about 15 \% contribution to the solar blend and the unidentified line makes 85 \% contribution, we now discard the Ba\,{\sc i} resonance line as a potential indicator of the Ba abundance.

\subsection{Ba\,{\sc ii} line abundances}
We have synthesized the Ba {\scs II} 5853 \AA\ lines in the solar twins using the ODFNEW grid of Kurucz model atmospheres \citet{caskur04} interpolated linearly to the stellar parameters (T$_{\rm eff}$, log~$g$, $\xi_{t}$) and metallicity presented in Table \ref{abu_solartwins}. The line list adopted in the spectrum synthesis is taken from \citep{mcwilliam1998} and the same was employed previously for the abundance analysis of dwarfs in stellar associations \citep{redlam15}. The synthesis has been done in a standard way using the {\it synth} driver of {\scs \bf MOOG} assuming LTE. Our adopted line list for Ba {\scs II} line includes the hyperfine structure components and isotopic shifts, and the fractional contribution of each isotope \citep{lodders03} to solar system Ba abundances. The total error on the measured barium abundances is estimated by adding in quadrature the sum of errors introduced by uncertainties in atmospheric parameters and spectrum synthesis. The total error in the meaured Ba abundances is about 0.03 dex.

\begin{figure}
\begin{center}
\includegraphics[trim=0.5cm 10.55cm 12.0cm 4.4cm, clip=true,height=0.24\textheight,width=0.38\textheight]{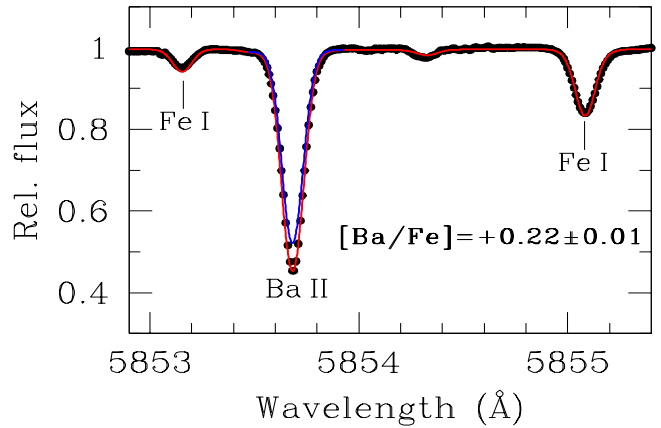}
\caption[]{ Comparison of synthetic profiles (red and blue) with the observed spectrum (black dots) of the solar analogue HD 96116 near the Ba {\scs II} 5853.7 \AA\ line. The best fit value of $+$0.22 dex and solar value of [Ba/Fe] are shown in red and blue, respectively.  }
\label{synth_hd96116ba}
\end{center}
\end{figure}

Our Ba abundance is derived assuming LTE. Non-LTE computations by \cite{korotin15} for the Ba$^+$ ion show that the differential non-LTE corrections in [Ba/Fe] are less than $-$0.01 dex for dwarfs covering the T$_{\rm eff}$ and gravity ranges of the solar twins considered here. Hence, the non-LTE Ba abundance of the Ba {\scs II} 5853 \AA\ line in stars will not be very different from the LTE value. We have also used other strong Ba {\scs II} 4554 \AA\ and 6496 \AA\ lines but the Ba {\scs II} abundance in this paper refers to Ba {\scs II} 5853 \AA\ line abundance, as the 5853 \AA\ line, the weakest of Ba {\scs II} lines, is widely used as an indicator of Ba abundance in stars. The non-LTE corrections for both Fe {\scs I} and Fe {\scs II} lines are negligible for the dwarfs over the [Fe/H] range covered by the present sample of solar analogues \citep{thevenin99}.
Therefore, the global behaviour of [Ba/Fe] (or any [X/Fe]) is unchanged with [Fe/H].

\begin{table*}
{\fontsize{8}{9}\selectfont
\caption{The barium abundances derived from the Ba {\scs I} and Ba {\scs II} lines in the solar twins along with the chromospheric activity index (log\,R$^{\prime}_{5876}$) determined from the line equivalent widths (EW$_{5876}^{(D_{3})}$) of the He {\scs I} 5875.6 \AA\ line.  } 
\vspace{0.2cm}
\label{abu_solartwins}
\begin{tabular}{cccccccccccccc}   \hline
 \multicolumn{1}{c}{HD} & \multicolumn{1}{c}{T$_{\rm eff}$ } & \multicolumn{1}{c}{$\log~g$} &\multicolumn{1}{c}{[Fe/H]} &
 \multicolumn{1}{c}{ $\xi_{t}$} & \multicolumn{1}{c}{Age} & \multicolumn{1}{c}{$v$~sin$i$} & \multicolumn{1}{c}{V$_{m}$} \vline & \multicolumn{3}{c}{[Ba {\scs II}/Fe] (dex) } & \multicolumn{1}{c}{log\,R$^{\prime}_{\rm HK}$ } & \multicolumn{1}{c}{log\,R$^{\prime}_{5876}$ } & \multicolumn{1}{c}{EW$_{5876}^{(He\,{\scs I}\,D_{3})}$} \\ \cline{7-8} \cline{9-11}
 \multicolumn{1}{l}{ } & \multicolumn{1}{c}{(K)} & \multicolumn{1}{c}{(dex)} &\multicolumn{1}{c}{(dex)} & \multicolumn{1}{c}{(km s$^{-1}$)} & \multicolumn{1}{c}{(Gyr)} & \multicolumn{2}{c}{(km s$^{-1}$)} \vline & \multicolumn{1}{c}{ $\lambda$ 5853 \AA\  } & \multicolumn{1}{c}{ $\lambda$ 4554 \AA\ } & \multicolumn{1}{c}{ $\lambda$ 6496 \AA\ } & \multicolumn{1}{c}{ } & \multicolumn{1}{c}{ } & \multicolumn{1}{c}{ (m\AA)}  \\  \hline
 
  2071$^{a}$ & 5724 & 4.49 & $-$0.08 &  0.96 &   3.5$\pm$0.8 &   1.3 &   3.1 &  +0.12 & +0.12 & +0.14 & -4.931 & -4.946 &  3.76 \\
  8406$^{a}$ & 5730 & 4.48 & $-$0.10 &  0.95 &   4.1$\pm$0.8 &   1.6 &   2.8 &  +0.13 & +0.12 & +0.10 & -4.672 & -4.742 &  9.36 \\
     20782 &  5776 &  4.34 & $-$0.06 &  1.04 &   8.1$\pm$0.4 &   1.3 &   3.1 &  +0.01 & +0.02 & +0.02 & -4.924$\pm$0.041 & -4.942 &  3.85 \\
     27063 &  5779 &  4.47 & $+$0.06 &  0.99 &   2.8$\pm$0.6 &   1.6 &   3.1 &  +0.09 & +0.10 & +0.09 & -4.749$\pm$0.077 & -4.647 & 13.30 \\
     28471 &  5754 &  4.38 & $-$0.05 &  1.02 &   7.3$\pm$0.4 &   1.3 &   3.1 &   0.00 &  0.00 & +0.01 & -5.010$\pm$0.043 & -4.993 &  2.76 \\
     38277 &  5860 &  4.27 & $-$0.07 &  1.17 &   7.8$\pm$0.4 &   1.4 &   3.2 &  +0.02 & +0.01 & +0.04 & -5.024$\pm$0.015 & -5.068 &  1.40 \\
     45184 &  5871 &  4.45 & $+$0.05 &  1.06 &   2.7$\pm$0.5 &   1.4 &   3.5 &  +0.07 & +0.07 & +0.05 & -4.896$\pm$0.073 & -4.905 &  4.72 \\
     45289 &  5718 &  4.28 & $-$0.02 &  1.06 &   9.4$\pm$0.4 &   1.3 &   3.2 &  -0.05 & -0.05 & -0.06 & -5.033$\pm$0.020 & -5.050 &  1.71 \\
     71334 &  5701 &  4.37 & $-$0.07 &  0.98 &   8.8$\pm$0.4 &   1.2 &   3.3 &  -0.02 &  0.00 & +0.01 & -4.995$\pm$0.019 & -4.998 &  2.68 \\
     78429 &  5756 &  4.27 & $+$0.08 &  1.05 &   8.3$\pm$0.4 &   1.4 &   3.4 &  -0.04 & -0.06 & -0.05 & -4.921$\pm$0.076 & -4.931 &  4.09 \\
     88084 &  5768 &  4.42 & $-$0.09 &  1.02 &   5.9$\pm$0.6 &   1.2 &   2.8 &   0.00 & +0.01 & +0.02 & -4.982$\pm$0.016 & -4.903 &  4.75 \\
     92719 &  5813 &  4.49 & $-$0.11 &  1.00 &   2.5$\pm$0.6 &   1.6 &   3.3 &  +0.15 & +0.14 & +0.14 & -4.860$\pm$0.067 & -4.800 &  7.60 \\
  96116$^{a}$& 5846 & 4.50 & $+$0.01 &  1.02 &   0.7$\pm$0.7 &   1.7 &   3.2 &  +0.22 & +0.18 & +0.19 & -4.719 & -4.614 & 14.80 \\
     96423 &  5714 &  4.36 & $+$0.11 &  0.99 &   7.3$\pm$0.6 &   1.2 &   3.4 &  -0.03 & -0.02 & -0.05 & -5.021$\pm$0.019 & -5.045 &  1.80 \\
    134664 &  5853 &  4.45 & $+$0.09 &  1.01 &   2.4$\pm$0.5 &   1.4 &   3.0 &  +0.07 & +0.07 & +0.08 & -4.871$\pm$0.082 & -4.760 &  8.90 \\
    146233 &  5809 &  4.43 & $+$0.05 &  1.02 &   4.0$\pm$0.5 &   1.5 &   3.5 &  +0.06 & +0.03 & +0.04 & -4.923$\pm$0.079 & -4.841 &  6.40 \\
    183658 &  5809 &  4.40 & $+$0.04 &  1.04 &   5.2$\pm$0.5 &   1.3 &   3.3 &  -0.03 & -0.04 & -0.05 & -4.983$\pm$0.018 & -5.028 &  2.10 \\
 208704$^{a}$& 5828 & 4.35 & $-$0.09 &  1.08 &   7.4$\pm$0.4 &   1.4 &   3.2 &  -0.01 & 0.00 & 0.00 & -4.975 & -4.920 &  4.35 \\
    210918 &  5748 &  4.32 & $-$0.09 &  1.07 &   9.1$\pm$0.4 &   1.3 &   2.0 &  +0.01 & 0.00 & 0.00 & -5.010$\pm$0.025 & -5.061 &  1.52 \\
    220507 &  5690 &  4.25 & $+$0.01 &  1.07 &   9.8$\pm$0.4 &   1.4 &   3.3 &  -0.05 & -0.07 & -0.08 & -5.050$\pm$0.018 & -5.074 &  1.30 \\
 222582$^{a}$ &  5784 &  4.36 &\,0.00 &  1.07 &   7.0$\pm$0.4 &   1.3 &  3.2 &  -0.03 & -0.04 & -0.04 & -4.978 & -5.040 &  1.89 \\
    59967$^{a}$ & 5847 & 4.54 & $-$0.02 &  1.17 & 0.4$\pm$0.4 &  3.7 &  3.1 &  +0.33 & +0.31 & +0.31 & -4.396 & -4.389 & 28.80 \\
   202628$^{a}$ & 5833 & 4.54 & \,0.00&  0.99 & 0.4$\pm$0.4 &  2.6 &  3.2 &    +0.23 & +0.24 & +0.24 & -4.720 & -4.667 & 12.40 \\
    42807$^{a}$ & 5737 & 4.49 & $-$0.02 &  1.12 & 2.9$\pm$1.0 &  3.8 &  2.9 &  +0.31 & +0.29 & +0.28 & -4.445 & -4.432 & 25.50 \\

\hline
\end{tabular}
\flushleft{ 
 Note -- log\,R$^{\prime}_{\rm HK}$ indices are taken from \cite{lovis11} for all but the stars marked with symbol ``$^{a}$'' whose values are drawn from \cite{ramirez14}. We assume a conservative uncertainty of about $\pm$0.08 in log\,R$^{\prime}_{\rm HK}$ values drawn from \cite{ramirez14}. The typical uncertainties in [Ba/Fe], log\,R$^{\prime}_{5876}$ and the EW of He {\scs I} D$_{3}$ profiles are 0.03 dex, 0.06 dex and 2 m\AA, respectively. }  }
\end{table*}

An example of synthetic spectra fit to the observed spectrum of HD 96116 near the Ba {\scs II} line is shown in the Figure \ref{synth_hd96116ba}. Results for the abundance ratios [X/Fe] of Ba {\scs II} lines in the solar analogues are provided in the Table \ref{abu_solartwins} along with the chromospheric activity indices and EWs of the He {\scs I} 5875.6 \AA\ line whose details will be given in later sections.

\begin{figure}
\begin{center}
\includegraphics[trim=0.1cm 9.4cm 8.0cm 4.2cm, clip=true,height=0.25\textheight,width=0.55\textheight]{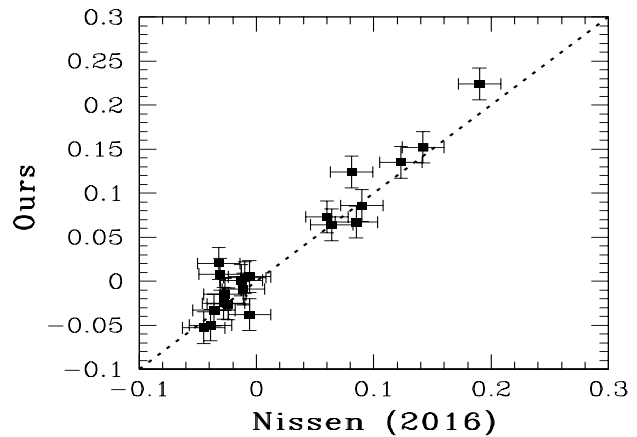}
\caption[]{Comparison of our [Ba/Fe] ratios from the Ba {\scs II} 5853 \AA\ line in solar twins with those from \cite{nissen16}.} 
\label{nissen16}
\end{center}
\end{figure}

A comparison of our [Ba/Fe] from the Ba {\scs II} 5853 \AA\ line measured using the Kurucz grid of LTE models with Ba abundances derived using the MARCS models by \cite{nissen16} is offered in Figure \ref{nissen16} for the 21 stars in common between the studies. Figure \ref{nissen16} highlights satisfactory agreement of measured abundances between the analyses based on different model photosphere grids and methods. The mean difference (ours-Nissen) of 0.008$\pm$0.021 dex in the barium abundance ratios between the analyses is pleasing.   

A key variable in understanding the barium puzzle is the age of a star. Stellar ages may be estimated in several ways. The fundamental ('thermodynamic') estimate comes from fitting basic stellar parameters to theoretical isochrones. Several alternative and empirical methods have been calibrated. One of these uses the surface lithium abundance and is anticipated to be of value when considering, as here, a sample of very similar stars. After discussing the ages from isochrones, we consider the Li abundance -- age relation.

The ages for the present sample solar twins were derived in \cite{nissen15,ramirez14} using the precise values of spectroscopic T$_{\rm eff}$ and log~${g}$, where the stars were compared to a set of isochrones in the T$_{\rm eff}-$log~${g}$ diagram. The Aarhus Stellar Evolution Code (ASTEC) described by \cite{christensen08} has been used by \cite{nissen16} to calculate isochrones in the T$_{\rm eff}-$log~${g}$ plane for a range of compositions covering those of the solar twins. Stellar ages with internal errors less than 1 Gyr were obtained in \cite{nissen16} using precise values of the observed quantities T$_{\rm eff}$, log~${g}$, [Fe/H] and [$\alpha$/Fe]. Ages for the stars HD 59967, HD 202628 and HD 42807 taken from \cite{ramirez14} have been estimated by comparing the location of stars in the T$_{\rm eff}-$log~${g}$ with a set of Yonsei-Yale (YY) isochrones \citep{yi01,kim02}. Given a set of observed quantities (T$_{\rm eff}$, log~${g}$, [Fe/H]) and their associated errors, YY isochrones were generated with a fine spacing of 0.02 dex, 0.005 in [Fe/H] and stellar mass, respectively. The final stellar age and 1$\sigma$ error is determined by adopting a probabilistic approach \cite{ramirez13,ramirez14}, where the probability that a given isochrone point represent well the observed quantities within the associated uncertainties. However, as explained in \cite{nissen16}, the ages derived from the YY isochrones differ slightly from those constrained from the ASTEC models. In order to place the ages on a common scale, we have transformed ages of the three stars from \cite{ramirez14} to \cite{nissen16}'s scale using the linear relation connecting the ASTEC and YY isochrone ages (equation 1 in \cite{nissen16}). Such a transformation reduced the ages of 0.6 Gyr each for HD 59967 and HD 202628 from \cite{ramirez14} to 0.4 Gyr each while the age of HD 42807 (2.8 Gyr) increased by 0.1 Gyr. The ages on the ASTEC scale provided in Table \ref{abu_solartwins} run from 400 Myr to 10 Gyr. The lithium abundance in solar twins may provide an alternative indicator of age with the older stars having lower Li content than the younger ones \citep{tucci15,carlos16}.

\begin{table}
{\fontsize{8}{9}\selectfont
\caption{The derived Li abundances of solar twins in comparison with previous analyses.} 
\vspace{0.2cm}
\label{li_abu}
\begin{tabular}{ccccc}   \hline
\multicolumn{1}{c}{HD} \vline & \multicolumn{3}{c}{log\,$\epsilon$(Li)LTE} \vline &\multicolumn{1}{c}{log\,$\epsilon$(Li)non-LTE}  \\ \cline{2-4}
\multicolumn{1}{c}{ }  \vline & \multicolumn{1}{c}{DM14} &\multicolumn{1}{c}{MC16} &\multicolumn{1}{c}{ours} \vline &\multicolumn{1}{c}{ours}   \\ 
\hline

   2071   &  1.38$\pm$0.07  &  1.39$\pm$0.03  &  1.38$\pm$0.03  &  1.42  \\
   8406   &  1.70$\pm$0.05  &  1.69$\pm$0.01  &  1.69$\pm$0.02  &  1.72  \\
 20782$^{a}$ &   $<$0.47    &  0.67$\pm$0.10  &  0.62$\pm$0.07  &  0.66  \\  
   27063  &   1.65$\pm$0.05 &  1.67$\pm$0.02  &  1.65$\pm$0.02  &  1.69  \\ 
   28471  &      $<$0.73    &  0.86$\pm$0.09  &  0.85$\pm$0.06  &  0.89  \\  
   38277  &   1.58$\pm$0.07 &  1.56$\pm$0.03  &  1.55$\pm$0.03  &  1.57  \\  
   45184  &   2.06$\pm$0.04 &  2.08$\pm$0.02  &  2.08$\pm$0.02  &  2.10  \\  
 45289$^{b}$ &   $<$0.47    &     $<$0.57     &  0.45$\pm$0.09  &  0.49  \\  
   71334  &      $<$0.65    &  0.58$\pm$0.12  &  0.55$\pm$0.07  &  0.59  \\  
   78429  &      $<$0.35    &  0.59$\pm$0.15  &  0.40$\pm$0.12  &  0.44  \\  
   88084  &      $<$1.10    &  0.96$\pm$0.09  &  0.95$\pm$0.08  &  0.99  \\  
   92719  &   1.90$\pm$0.04 &  1.88$\pm$0.01  &  1.88$\pm$0.02  &  1.90  \\  
   96116  &   2.12$\pm$0.07 &  2.19$\pm$0.01  &  2.15$\pm$0.02  &  2.17  \\  
   96423  &   1.88$\pm$0.05 &  1.88$\pm$0.01  &  1.87$\pm$0.02  &  1.92  \\  
   134664 &   2.10$\pm$0.05 &  2.08$\pm$0.02  &  2.08$\pm$0.02  &  2.11  \\  
   146233 &   1.57$\pm$0.07 &  1.58$\pm$0.02  &  1.55$\pm$0.03  &  1.59  \\  
   183658 &   1.09$\pm$0.10 &  1.24$\pm$0.06  &  1.05$\pm$0.03  &  1.09  \\  
   208704 &   1.09$\pm$0.10 &  1.05$\pm$0.04  &  1.06$\pm$0.03  &  1.09  \\  
   210918 &      $<$0.28    &  0.67$\pm$0.13  &  0.30$\pm$0.12  &  0.34  \\  
   220507 &      $<$0.56    &     $<$0.50     &  0.27$\pm$0.15  &  0.31  \\  
 222582$^{c}$ &   $<$0.45   &  0.89$\pm$0.07  &  0.88$\pm$0.06  &  0.92  \\  
    59967 &     $\dots$     &     $\dots$     &  2.74$\pm$0.02  &  2.72  \\
   202628 &     $\dots$     &     $\dots$     &  2.23$\pm$0.02  &  2.25  \\
    42807 &     $\dots$     &     $\dots$     &  2.02$\pm$0.03  &  2.05  \\

\hline
\end{tabular} }
\flushleft{ 
 Ref -- DM14: \cite{delgado14}; MC16: \cite{carlos16}\\
 Detected planets with -- $^{(a)}$ 1.8 M$_{Jup}$ \citep{jones06}; $^{(b)}$ 0.04 M$_{Jup}$ \citep{mayor11}; $^{(c)}$ 7.8 M$_{Jup}$ \citep{butler06}.
  }
\end{table}

\begin{figure}
\begin{center}
\includegraphics[trim=0.3cm 9.9cm 5.7cm 4.3cm, clip=true,height=0.26\textheight,width=0.55\textheight]{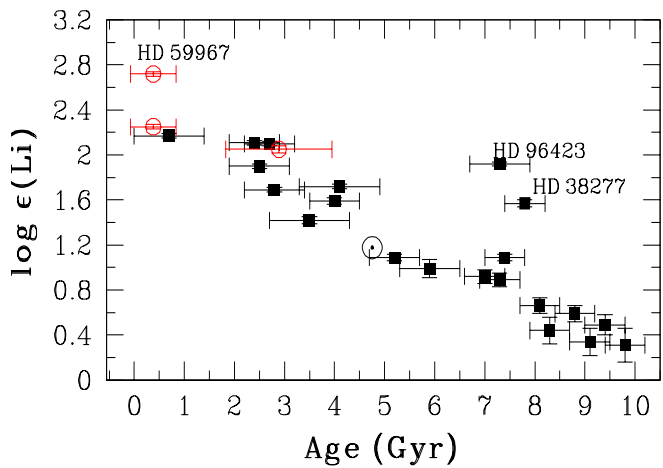}
\caption[]{ Connection between the non-LTE lithium abundances and stellar ages of the solar twins taken from \cite{nissen16} (squares) and \cite{ramirez14} (red circles). The Sun is designated with the symbol $\odot$. }
\label{li_age}
\end{center}
\end{figure}

\begin{figure}
\begin{center}
\includegraphics[trim=0.1cm 9.3cm 6.9cm 4.2cm, clip=true,width=0.55\textheight]{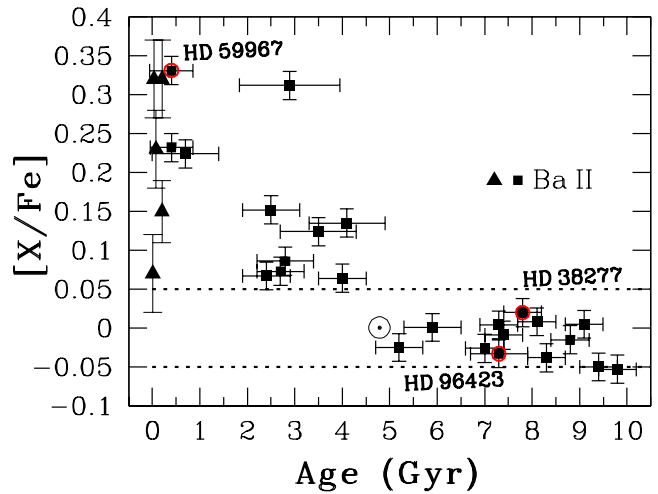}
\caption[]{[Ba/Fe] versus age for the solar twins (squares) in this study and for the stellar associations (triangles) drawn from \cite{redlam15}. The outliers (HD 38277, HD 59967, HD 96423) are also encirled with red circles and the Sun is shown with the symbol $\odot$. The horizontal dotted lines mark the bounds of $\pm$0.05 dex in [X/Fe]. }
\label{ba_solartwins}
\end{center}
\end{figure}

\begin{figure*}
\begin{center}
\includegraphics[trim=0.5cm 10.5cm 4.0cm 4.4cm, clip=true,height=0.26\textheight,width=0.85\textheight]{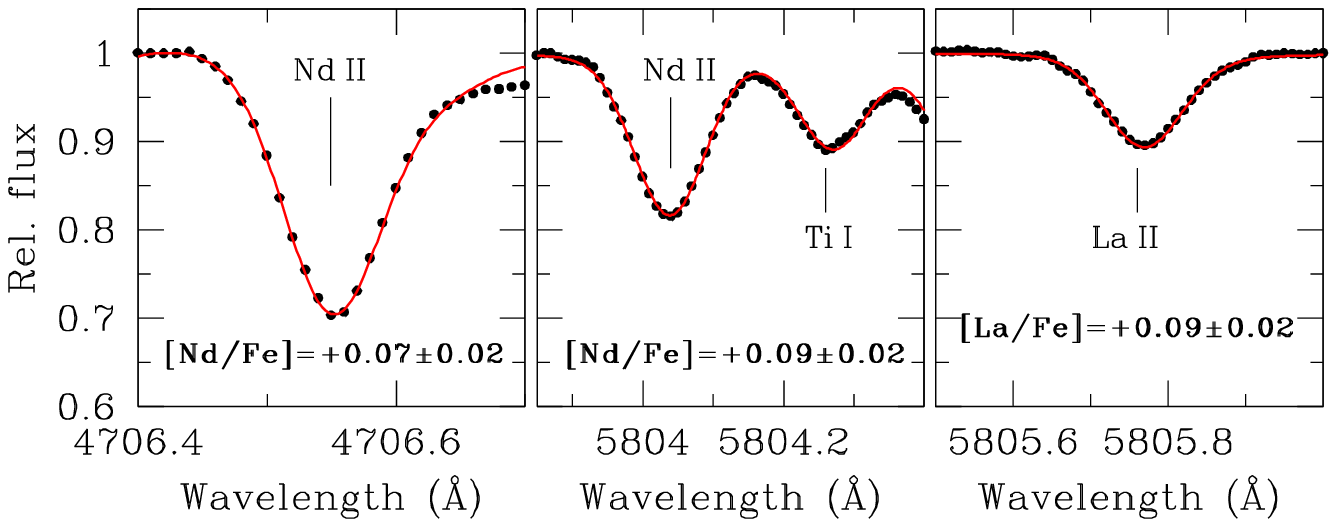}
\caption[]{ Comparison of synthetic profiles (red lines) with the observed spectrum (black dots) of the solar anaogue HD 96116 near the Nd {\scs II} and La {\scs II} lines. In each panel, the best fit synthetic profile and the corresponding abundance values of Nd and La relative to Fe are given.  }
\label{synth_hd96116nd}
\end{center}
\end{figure*}

\subsection{Evolution and Li age}

The lithium abundances for the sample of stars drawn from \cite{nissen16} have been already measured by \cite{carlos16}. To place the Li abundances of our complete sample of stars on a homogeneous scale, we have re-analysed the \cite{nissen16}'s sample for lithium abundances. Neglecting the possible $^{6}$Li contribution, we measured the LTE lithium abundance of solar twins via the spectrum synthesis of the Li 6707.8 \AA\ line adopting the hyperfine structure data from \cite{ghezzi09}. Then the non-LTE corrections were applied to LTE Li abundances based on the non-LTE calculations by \cite{lind09}. Results for both LTE and non-LTE lithium abundances are provided in Table \ref{li_abu} in comparison with previous estimates of Li in our stars which indicate a satisfactory agreement among the analyses.

The strong correlation of Li abundance with the stellar age shown in Figure \ref{li_age} is similar to the one discussed extensively in \cite{carlos16} (see, their Figure 6) where they showed Li abundances as a function of ages derived previously by \cite{nissen15} using YY isochrones. Two of the three stars whose primary age estimates are drawn from the analysis of \cite{ramirez14} integrate well into the decreasing trend of lithium with age. But the star HD 59967 from \cite{ramirez14} and two more stars (HD 96423 and HD 38277) from \cite{nissen16}'s sample turned out to be outliers with their Li abundances preferring young ages than the isochrone based ages \citep{ramirez14,nissen16}. Whereas the log\,R$^{\prime}_{\rm HK}$ indices of $-$5.02 each (Table \ref{abu_solartwins}) for HD 96423 and HD 38277 prefers an age much older than the Sun.

Note, however, the strong correlation of log\,$\epsilon$(Li) with age shown by \cite{carlos16} and us is not confirmed by \cite{thevenin17} whose selection of solar twins covering the same T$_{\rm eff}$, log~${g}$, [Fe/H] and mass range as \cite{carlos16} is drawn from Li abundances in a large sample of stars by \cite{delgado14}. \citeauthor{thevenin17}'s mean trend of Li with age follows that in Figure \ref{li_age} but shows a large scatter in Li abundance at all ages. \cite{thevenin17} argue that the lithium abundance is sensitive to various physical conditions during the early pre-main sequence stage, for example, different overshooting values. Therefore, the Li abundance in a solar twin is not solely dependent on a star's age. The isochrone based ages derived using the precise values of T$_{\rm eff}$, log~${g}$ and [Fe/H] provide a robust measure of ages of solar twins. In the following analysis, we adopt the isochrone based ages in Table \ref{abu_solartwins} and our measured values of [Ba/Fe] to investigate the variation of [Ba/Fe] with stellar age. We stress that our major conclusions, as we see in the following sections, will not be influenced by the three outliers mentioned previously. 

Entries for Ba {\scs II} in Table \ref{abu_solartwins} are plotted in Figure \ref{ba_solartwins} against the isochrone age. This shows, as indicated in \citep{nissen16}, that the [Ba {\scs II}/Fe] values of the solar twins exhibit an increasing trend and increasing scatter with decreasing stellar age. For stars older than 5 Gyr, [Ba/Fe] flattens out with a spread of $\pm$ 0.05 dex about the mean trend of [Ba/Fe] $\sim -0.02$ dex. For stars younger than 5 Gyr, the typical scatter of $\pm$0.05 dex observed for the older stars evolves with age to maximum value. Based on the lithium abundances (Figure \ref{li_age}), adopting younger ages for the outliers will amplify the scatter in [Ba {\scs II}/Fe] with age for the solar twins younger than the Sun. The Figure \ref{abu_solartwins} also display the mean [Ba/Fe] values with a scatter of 0.05 dex measured for stellar associations by \cite{redlam15}.

\section{Neutron-capture elements}
The abundances of the heavy elements La, Ce, Nd and Sm in solar twins have been measured here differentially, line-by-line, relative to the Sun adopting the line list from \cite{redlam15}. A representative example of synthetic spectra fit to the observed spectrum of HD 96116 near the La {\scs II} and Nd  {\scs II} lines is shown in Figure \ref{synth_hd96116nd} for the abundances listed in Table \ref{abu_heavy_solartwins}. Results for the abundance ratios [X/Fe] of the heavy $s$-process elements are given in the Table \ref{abu_heavy_solartwins}.
  
\begin{table}
{\fontsize{8}{9}\selectfont
\caption{Abundance ratios ([X/Fe]) for the heavy $s$-process elements X$=$La, Ce, Nd and Sm derived from the spectra of solar twins.} 
\vspace{0.2cm}
\label{abu_heavy_solartwins}
\begin{tabular}{ccccc}   \hline
\multicolumn{1}{l}{HD} &\multicolumn{1}{c}{[La/Fe]} &\multicolumn{1}{c}{[Ce/Fe]} &\multicolumn{1}{c}{[Nd/Fe]} & \multicolumn{1}{c}{[Sm/Fe]} \\
\multicolumn{1}{l}{ } &\multicolumn{1}{c}{(dex)} &\multicolumn{1}{c}{(dex)} &\multicolumn{1}{c}{(dex)} & \multicolumn{1}{c}{(dex)} \\ \hline
 
      2071 &  0.06$\pm$0.01 &  0.06$\pm$0.01 &  0.05$\pm$0.02 &  0.04$\pm$0.02  \\
      8406 &  0.05$\pm$0.01 &  0.05$\pm$0.01 &  0.06$\pm$0.02 &  0.03$\pm$0.02  \\
     20782 & -0.00$\pm$0.02 &  0.01$\pm$0.01 & -0.01$\pm$0.02 &  0.01$\pm$0.02  \\
     27063 &  0.02$\pm$0.02 &  0.01$\pm$0.01 &  0.01$\pm$0.02 &  0.02$\pm$0.02  \\
     28471 &  0.02$\pm$0.02 &  0.02$\pm$0.02 &  0.01$\pm$0.02 &  0.02$\pm$0.02  \\
     38277 & -0.02$\pm$0.02 & -0.03$\pm$0.01 &  0.00$\pm$0.02 &  0.01$\pm$0.02  \\
     45184 &  0.01$\pm$0.02 &  0.01$\pm$0.01 &  0.03$\pm$0.02 &  0.05$\pm$0.02  \\
     45289 & -0.04$\pm$0.02 & -0.05$\pm$0.01 & -0.06$\pm$0.02 & -0.04$\pm$0.01  \\
     71334 &  0.01$\pm$0.02 &  0.00$\pm$0.02 & -0.01$\pm$0.01 &  0.01$\pm$0.02  \\
     78429 & -0.02$\pm$0.01 & -0.05$\pm$0.01 & -0.04$\pm$0.01 & -0.03$\pm$0.01  \\
     88084 &  0.01$\pm$0.01 &  0.02$\pm$0.02 &  0.03$\pm$0.02 &  0.02$\pm$0.02  \\
     92719 &  0.07$\pm$0.01 &  0.06$\pm$0.02 &  0.04$\pm$0.02 &  0.06$\pm$0.02  \\
     96116 &  0.08$\pm$0.02 &  0.06$\pm$0.02 &  0.07$\pm$0.02 &  0.05$\pm$0.02  \\
     96423 & -0.02$\pm$0.01 & -0.01$\pm$0.02 & -0.02$\pm$0.01 & -0.02$\pm$0.02  \\
    134664 &  0.02$\pm$0.02 &  0.03$\pm$0.01 &  0.05$\pm$0.02 &  0.01$\pm$0.02  \\
    146233 &  0.04$\pm$0.02 &  0.04$\pm$0.02 &  0.05$\pm$0.01 &  0.01$\pm$0.01  \\
    183658 & -0.01$\pm$0.01 &  0.01$\pm$0.01 &  0.05$\pm$0.01 &  0.03$\pm$0.02  \\
    208704 & -0.02$\pm$0.02 & -0.00$\pm$0.02 & -0.01$\pm$0.01 &  0.01$\pm$0.02  \\
    210918 & -0.01$\pm$0.02 &  0.00$\pm$0.02 & -0.01$\pm$0.02 &  0.00$\pm$0.02  \\
    220507 & -0.02$\pm$0.01 & -0.04$\pm$0.02 & -0.03$\pm$0.02 & -0.06$\pm$0.01  \\
    222582 & -0.01$\pm$0.01 & -0.04$\pm$0.01 & -0.06$\pm$0.01 & -0.06$\pm$0.02  \\
     59967 & +0.06$\pm$0.02 & +0.06$\pm$0.02 & +0.05$\pm$0.02 & +0.05$\pm$0.02  \\
    202628 & +0.05$\pm$0.02 & +0.05$\pm$0.02 & +0.06$\pm$0.02 & +0.05$\pm$0.02  \\
     42807 & +0.04$\pm$0.02 & +0.05$\pm$0.02 & +0.06$\pm$0.02 & +0.06$\pm$0.02  \\
\hline
\end{tabular} }
\end{table}

\begin{figure}
\begin{center}
\includegraphics[trim=0.1cm 9.3cm 6.7cm 4.2cm, clip=true,width=0.54\textheight]{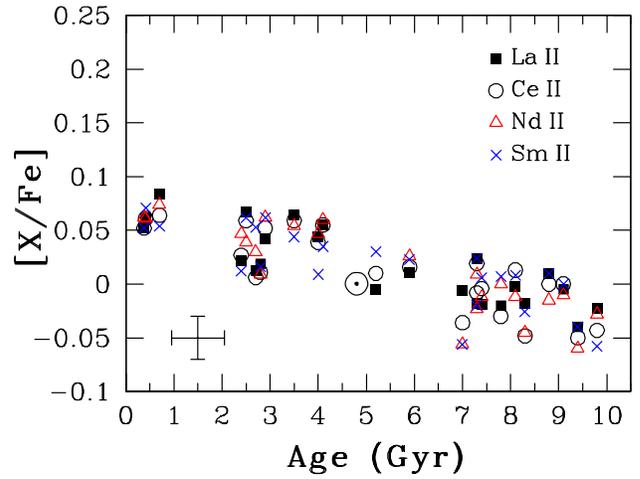}
\caption[]{[X/Fe] abundance ratios for an element X$=$La, Ce, Nd and Sm in solar twins as a function of stellar age. The Sun is shown with the symbol $\odot$ and the typical error bars in [X/Fe] and age are also shown. }
\label{heavy_solartwins}
\end{center}
\end{figure}

We probe through Figure \ref{heavy_solartwins} the variation of abundance ratios [X/Fe] with stellar age for the suite of heavy elements La, Ce Nd and Sm in solar twins. Figure \ref{heavy_solartwins} suggests that the [X/Fe] values measured for the heavy $s$-process elements display a smoothly increasing weak trend with decreasing stellar age. The typical spread of $\pm$0.05 dex in the [X/Fe] for La$-$Sm distribution at any given stellar age is comparable to the very similar scatter found for the heavy element Ba (Figure \ref{ba_solartwins}) but for stars older than the Sun (ages$>$5 Gyr). But, the stars having larger values of [Ba {\scs II}/Fe] are not similarly enriched in other heavy elements -- La, Ce, Nd, and Sm -- whose abundance is expected to be dominated by the $s$-process which also controls the Ba abundance. For stars younger than the Sun (ages$<$5 Gyr), the correlation of [Ba/Fe] with age is very weak and dominated by the large scatter at any given stellar age (Figure \ref{ba_solartwins}) unlike the clean trend with little spread seen for the [La/Fe] et al. abundances.

There exists (i) striking differences between the [Ba {\scs II}/Fe] values and [La/Fe] et al. abundances in young stars (ages$<$5 Gyr) but not in stars older than the Sun (ages$>$5 Gyr) and (ii) the [La/Fe] et al. correlate well with stellar age and the scatter at a given age is much cleaner than for [Ba\,{\scs II}/Fe] versus age. The magnitude of [Ba {\scs II}/(La-Sm)] values range over 0.25 dex at very young ages among these solar-type stars; this range is almost ten times larger than the $\sigma$ in [X/Fe] ratios from a given star. Such differences among the heavy $s$-process elements are puzzling in the context of LTE line formation in stellar photospheres and the $s$-process operation in AGB stars. 

It is a fundamental aspect of $s$-process operation that the higher Ba yield should be accompanied by higher yields of especially La and Ce, also Nd, and Sm \citep{busso99} which is not satisfied by elements La to Sm relative to Ba {\scs II} in young stars. Contributions of the $s$-process to solar abundances run from 85\% for Ba, 75\% for La, 81\% for Ce, 47\% for Nd and 34\% for Sm \citep{burris2000}. Operation of the $s$-process under different conditions will have very little effect on the relative yields of elements Ba to Sm but may change the Ba to Y-Zr ratios depending on the strength of the neutron flux. 

\begin{figure}
\begin{center}
\includegraphics[trim=0.1cm 9.3cm 6.7cm 4.2cm, clip=true,height=0.28\textheight,width=0.74\textwidth]{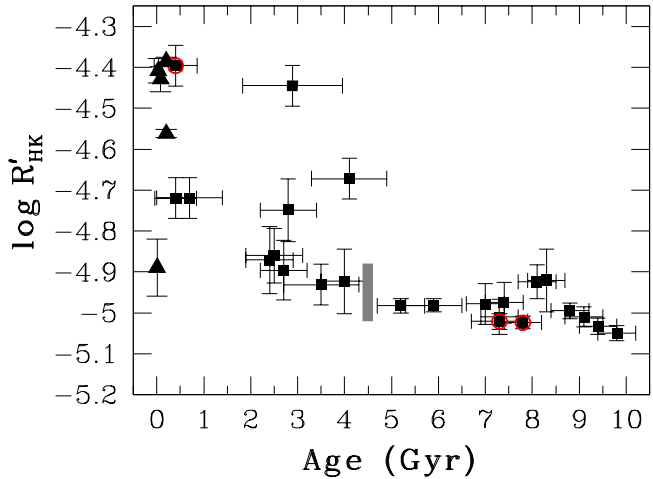}
\caption[]{Stellar chromospheric activity index as a function of ages of the solar twins (filled squares) studied here and the young associations (filled triangles) from \cite{redlam15}. The outliers, as in Figure \ref{ba_solartwins}, are encirled with red circles. The shaded gray bar at 4.5 Gyr represents the range in activity covered by the 11-year solar cycle.  }
\label{rhk_age}
\end{center}
\end{figure}

Therefore, results for the heavy elements from Table \ref{abu_solartwins}, Table \ref{abu_heavy_solartwins}, Figure \ref{ba_solartwins} and Figure \ref{heavy_solartwins} suggest that the solution to the large deviation of the Ba {\scs II} line abundance relative to La to Sm may be the same as the one controlled by stellar phenomena not accounted for in LTE abundance analyses, e.g., stellar activity is severe at very young ages but decays slowly as the star ages. The parameters of most relevance to young stars may be the rapid rotation, strong magnetic fields and, thus, their active chromospheres. In the next section, we explore the dependence of heavy element abundances on chromospheric activity levels in the solar twins.

\section{Chromospheric activity}
As noted in the introduction, the sample of solar twins in this study was analysed previously in \cite{nissen15,nissen16}. In this section, we explore for the first time a possible relation between abundance ratios [X/Fe] and stellar activity of the solar twins.

\begin{figure}
\begin{center}
\includegraphics[trim=0.6cm 9.6cm 6.7cm 4.5cm, clip=true,height=0.24\textheight,width=0.65\textwidth]{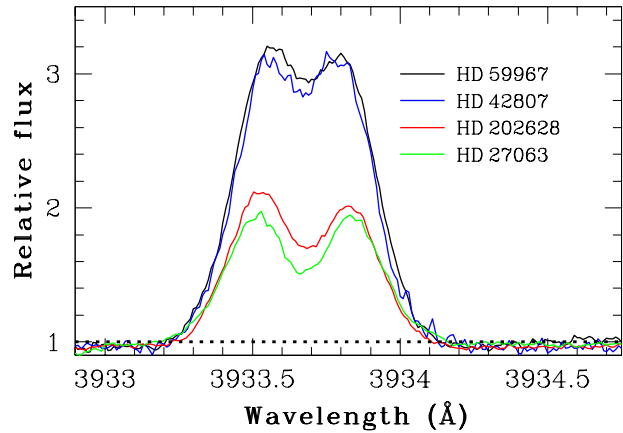}
\caption[]{ Comparison of Ca {\scs II} K line emission cores in the spectra of HD 59967, HD 42807, HD 202628 and HD 27063. }
\label{cak_twins}
\end{center}
\end{figure}

The chromospheric activity indices R$^{\prime}_{\rm HK}$ \citep[see, for definition,][]{lovis11} for all but the solar twins HD 2071, HD 8406, HD 42807, HD 59967, HD 96116, HD 202628, HD 208704, HD 222582 in Table \ref{abu_solartwins} are extracted from \cite{lovis11} who exploited multi-epoch HARPS spectra and converted the measured Ca {\scs II} H\&K emission line fluxes into R$^{\prime}_{\rm HK}$ indices using an updated calibration taking into account stellar metallicity. For the stars not considered by \cite{lovis11}, the R$^{\prime}_{\rm HK}$ indices are taken from the measurements of \cite{ramirez14} whose R$^{\prime}_{\rm HK}$ values for stars in common with \cite{lovis11} are in excellent agreement and on average only 0.005$\pm$0.025 higher. Therefore, our extracted R$^{\prime}_{\rm HK}$ values are unaffected by systematic offsets as both the studies have activity indices measured using the common methods (see Figure \ref{cak_twins}).

A sanity check on the reliability of the log\,R$^{\prime}_{\rm HK}$ values of stars taken from the measurements of \cite{ramirez14} is provided in Figure \ref{cak_twins} via a direct comparison of the Ca {\scs II} K emission line cores in the spectra of HD 59967, HD 42807, HD 202628. In addition, we compare the Ca {\scs II} K profile of HD 202628 with a star HD 27063 whose log\,R$^{\prime}_{\rm HK}$ index ($-$4.75$\pm$0.08) measured from the multi-epoch spectra by \cite{lovis11} is very similar to log\,R$^{\prime}_{\rm HK}$ of $-$4.72 measured for HD 202628 in \cite{ramirez14}. It is obvious from the Figure \ref{cak_twins} and log\,R$^{\prime}_{\rm HK}$ values (Table \ref{abu_solartwins}) that the log\,R$^{\prime}_{\rm HK}$ values derived by \cite{ramirez14} are as reliable as those measured using the multi-epoch HARPS of solar twins by \cite{lovis11}. Also confirmed by the Figure \ref{cak_twins} are the very similar and much higher levels of chromospheric activity for HD 59967 and HD 42807 than HD 202628.

However, we have cross-checked the determinations of R$^{\prime}_{\rm HK}$ indices with the chromospheric activity index derived from the equivalent width of the He {\scs I} 5876 \AA\ D$_{3}$ line. The D$_{3}$ line, as in the Sun, appears in absorption and is produced in the plage regions (bright regions) of the chromosphere \citep{landman81} and, thus, used as a diagnostic of chromospheric activity in late F through G to early K dwarfs \citep{danks85,wolff86}. 

Acknowledgment of a tight correlation of the EW of stellar D$_{3}$ absorption line with other activity indicators such as the coronal X-ray luminosity (L$_{\rm x}$) and flux in the Ca {\scs II} H and K chromospheric emission cores (F$^{\prime}_{\rm HK}$ or R$^{\prime}_{\rm HK}$) is provided by \cite{danks85} who examined the D$_{3}$ line in the high-resolution spectra of 21 unevolved stars of spectral types F, G and K. The formation zone of the He {\scs I} 5876 \AA\ D$_{3}$ absorption corresponds to a temperature of $\sim$10,000 K and is located between the upper chromosphere and corona whereas the Ca {\scs II} H and K emission lines are produced in the low and medium chromosphere and the X-rays in the stellar corona. The correlation of the equivalent width of the He {\scs I} D$_{3}$ line with other activity indicators is evidence of direct proportionality of R$^{\prime}_{\rm HK}$, the EW of D$_{3}$ and L$_{\rm x}$ to the fractional area of the stellar disk covered by active regions \citep{danks85}.

\begin{figure}
\begin{center}
\includegraphics[trim=0.1cm 8.7cm 9.5cm 4.5cm, clip=true,height=0.33\textheight,width=0.74\textwidth]{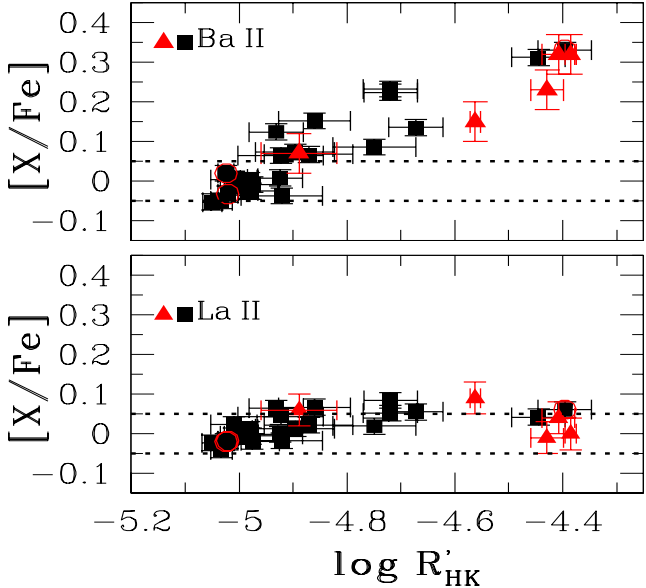}
\caption[]{[X/Fe] for the heavy elements Ba {\scs II} and La {\scs II} versus stellar chromospheric activity of the solar twins (squares) and the stellar associations (red triangles). The outliers, as in Figure \ref{rhk_age}, are encirled with red circles.  }
\label{rhk_ba}
\end{center}
\end{figure}

The He {\scs I} 5876 \AA\ D$_{3}$ line is free of severe blends and is easily measurable from the high-resolution and high S/N ratio spectra but is contaminated by a very weak telluric line. The spectra of the solar twins yield S/N ratios of 550 at the He {\scs I} D$_{3}$ line. The He {\scs I} D$_{3}$ line EWs were measured by fitting a Gaussian profile or by direct integration of the He {\scs I} line using the {\it splot} task in {\scs IRAF}. 
To convert the He {\scs I} D$_{3}$ line EWs into R$^{\prime}_{\rm HK}$ values, we derived a linear regression relation using the values of R$^{\prime}_{\rm HK}$ and EWs of D$_{3}$ lines (EW$_{5876}$\,(He\,{\scs I})) in Danks \& Lambert (1985; see their Figure 6) for their sample of F, G and K dwarfs. The empirical relation takes the form 

\begin{equation} \label{eq_ewvsrhk}
 EW_{5876}\,(He {\scs I}) = 8.49\,x\,-5.86 \, (r = 0.97)
\end{equation}
where $x=$ 10$^{5}\times$\,R$^{\prime}_{\rm HK}$ and r is the Pearson product-moment correlation coefficient. Our measured values of He {\scs I} D$_{3}$ line EWs and the chromospheric activity indices (R$^{\prime}_{5876}$) are given in Table \ref{abu_solartwins}. From Table \ref{abu_solartwins}, an excellent agreement of the chromospheric activity indices computed independently from the Ca {\scs II} H and K emission line fluxes and the EWs of He {\scs I} 5876 \AA\ D$_{3}$ lines is evident. The mean differences in activity indices (R$^{\prime}_{\rm HK}-$R$^{\prime}_{5876}$) of only $-$0.01$\pm$0.05 is comparable with typical errors in R$^{\prime}_{5876}$ of 0.06. We derived from the present spectra of the Sun the activity index of $-$4.93, a value falls in the log\,R$^{\prime}_{\rm HK}$ range $-$4.88 to $-$5.02 seen during the 11-year solar cycle \citep{hall09}. These results corroborate the reliability of the He {\scs I} 5876 \AA\ D$_{3}$ absorption line as a powerful diagnostic of stellar activity in FGK dwarfs.

We probe through the Figure \ref{rhk_age} and Figure \ref{rhk_ba} the variation of chromospheric activity with stellar age and the dependence of barium line abundances of solar twins on stellar activity, respectively. Stellar chromospheric activity is observationally known to decline steeply with stellar age for solar type field dwarfs of age less than the Sun and thereafter continues to decrease very slowly as stars get significantly older \citep{ramirez14}. Indeed, field stars are relatively old, rotate more slowly and have less chromospheric activity than their main sequence counterparts in OCs \citep{soderblom83,soderblom91,simon85,isaacson10}. As demonstrated in \cite{mamajek08}, the mean log\,R$^{\prime}_{\rm HK}$ values of OCs and associations derived from solar type dwarfs not only extend the R$^{\prime}_{\rm HK}$ versus age relation to very young ages of 5 Myr but have shown the striking decline of log\,R$^{\prime}_{\rm HK}$ by almost 1 dex between the T-Tauri epoch (1-10 Myr) and age of the Sun \citep[see Figure 6 in][]{mamajek08}. (This decline of chromospheric activity with age is sometimes used to provide an empirical method for dwarf stars.) 

\begin{figure*}
\begin{center}
\includegraphics[trim=0.6cm 9.0cm 6.1cm 4.2cm, clip=true,height=0.3\textheight,width=0.7\textheight]{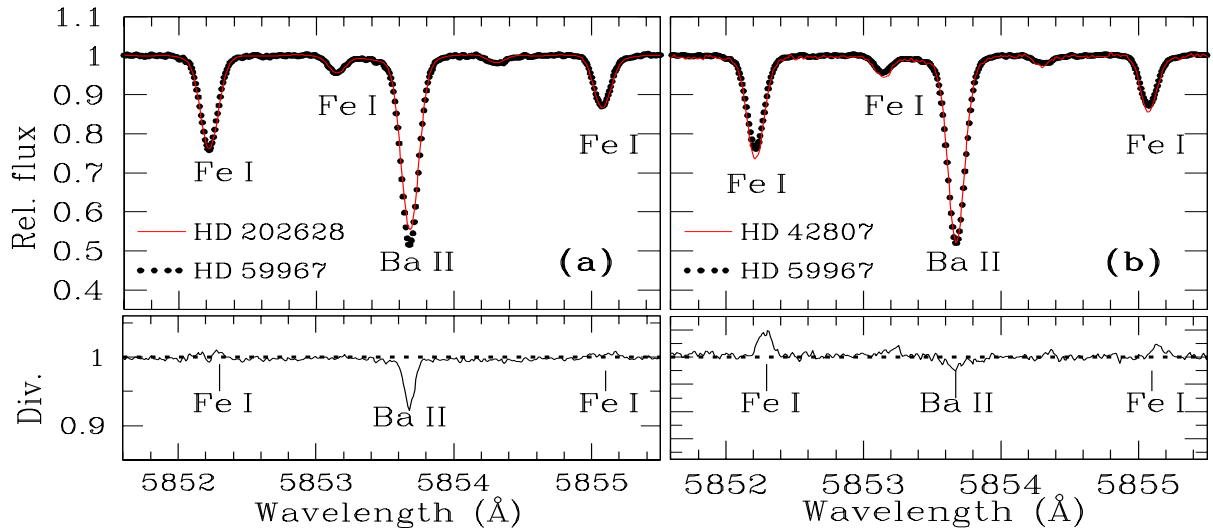}
\caption[]{Observed spectra of HD 59967 (dotted line) compared with HD 202628 and HD 42807 shown as continuous lines.  A panel whose y-axis reads Div. (division) is the residual flux resulted from the division of HD 59967 by HD 202628 and HD 42807, respectively. Panel (a) compares the solar twins of same age (0.4 Gyr) but with different levels of stellar activity: HD 59967 (log\,R$^{\prime}_{\rm HK}=$ $-$4.4) and HD 202628 (log\,R$^{\prime}_{\rm HK}=$ $-$4.7). Panel (b) compares stars with similar log\,R$^{\prime}_{\rm HK}$ indices but differ in ages: HD 59967 (0.4 Gyr) and HD 42807 (2.8 Gyr). A novel result arriving from this plot and the Figure \ref{rhk_ba} is that the strengthening of Ba {\scs II} 5853 \AA\ line (hence, overabundance of [Ba/Fe]) is associated with high levels of stellar activity but not with the stellar age (see text). }
\label{hd96116ba3}
\end{center}
\end{figure*}

From Figure \ref{rhk_age}, our sample of solar twins follows the declining trend of chromospheric activity with stellar age which resembles such variations shown for a large sample of solar twins in \cite{ramirez14}. The star-to-star dispersion at any given age is dominated by intrinsic changes in activity of the stars, where the stars younger than the Sun exhibit a wide range in activity of $-$4.4 to $-$5.0 \citep{ramirez14}. This is a pleasing result along with small variation in log\,R$^{\prime}_{\rm HK}$ indices for stars older than the Sun, compatible with the activity levels of the present-day solar cycle. The variation in the log\,R$^{\prime}_{\rm HK}$ index from about $-$4.88 to $-$5.02 of the Sun \citep{hall09} during the 11-year solar cycle is represented by a grey bar at 4.5 Gyr in Figure \ref{rhk_age}. 

The new result demonstrated in the Figure \ref{rhk_ba} is that the Ba {\scs II} abundance correlates well with the stellar chromospheric activity: The sample of solar twins yielding the Ba abundances very close to [La/Fe] et al. values are older and their chromospheres are less active than the Sun (log\,R$^{\prime}_{\rm HK}<$ $-$4.95). Among stars of high chromospheric activity (log\,R$^{\prime}_{\rm HK}>$ $-$4.95), the Ba abundance from the strong Ba {\scs II} lines increases with increasing activity. It is evident from Figure \ref{heavy_solartwins} and Figure \ref{rhk_ba} that the [X/Fe] for other heavy $s$-process elements such as La to Sm are almost unchanged with stellar activity.

Scatter in Figure \ref{rhk_ba} among [Ba/Fe] values at a given log\,R$^{\prime}_{\rm HK}$ is likely to come from the time-dependent chromospheric activity; the spectra providing the Ba abundances were not obtained at the same time as the R$^{\prime}_{\rm HK}$ were determined. For most stars, the R$^{\prime}_{\rm HK}$ was obtained from HARPS spectra at a time not so different from the exposures which were co-added and subjected to the abundance analysis. Thus, one is not surprised that the scatter in Figure \ref{rhk_ba} is appreciably less than that in Figure \ref{rhk_age} -- especially among the younger stars where chromospheric activity is more intense than the older stars. 

Striking evidence that the strengthening of the Ba {\scs II} line and, thus, the abundance of [Ba {\scs II}/Fe] in the solar twins is strongly associated with the stellar activity but weakly with stellar age is provided through the Figure \ref{hd96116ba3}. Figure \ref{hd96116ba3} offers a comparison of the observed spectra of (i) stars with the same age of 0.4 Gyr (HD 59967 and HD 202628) but with different levels of log\,R$^{\prime}_{\rm HK}$ values and (ii) stars with identical log\,R$^{\prime}_{\rm HK}$ values of $-$4.4 but with different ages (HD 59967 and HD 42807). Among stars with identical age (Panels (a) and (b)), the stronger Ba {\scs II} line is observed in HD 59967 with larger log\,R$^{\prime}_{\rm HK}$ value of $-$4.4 relative to $-$4.7 for HD 202628. Among stars with identical log\,R$^{\prime}_{\rm HK}$ values of $-$4.4 each (Panels (c) and (d)), the Ba {\scs II} lines have identical strengths (EWs) and are independent of stellar age. The scatter in Figure \ref{rhk_age} at ages of 3 Gyr and shorter is not reflected in Figure \ref{rhk_ba}. That is [X/Fe] for Ba {\scs II} correlates well with log\,R$^{\prime}_{\rm HK}$ and much better than with age (Figure \ref{ba_solartwins}). The linear regression relation between the [Ba/Fe] and log\,R$^{\prime}_{\rm HK}$ values for the 24 solar twins provide the Pearson product-moment correlation coefficient of $r$\,=\,0.92, meaning an almost perfect and strong correlation of [Ba/Fe] with chromospheric activity index. This simple illustration is at odds with reports of enhanced Ba nucleosynthesis with decreasing stellar (Galactic) age \citep{dorazi09,mishenina15} but provides a novel result that the abundance of [Ba/Fe] in young stars is strongly associated with high level of stellar activity and, thus, unrelated to aspects of the chemical evolution of stars.

The prime novel results of this paper -- (i) the trend of [Ba {\scs II}/Fe] with stellar activity and (ii) the higher abundance of Ba from the lines of Ba {\scs II} relative La$-$Sm for the youngest and chromospherically active solar analogues -- provide strong evidence that the large [Ba {\scs II}/Fe] values relative to La$-$Sm abundances noticed here at the very young ages where the stars have very active chromospheres are unrelated to aspects of stellar nucleosynthesis and Galactic chemical evolution. 

Model stellar atmospheres of the kind chosen here make no attempt to incorporate a chromosphere. The heating responsible for the chromosphere presumably raises the temperature in the photosphere-chromosphere interface above the temperatures predicted by present model stellar atmospheres.  The heating may also disturb the granulation pattern and associated velocity field and so change the microturbulence. Although the Ba {\scs II} 5853 \AA\ line is insensitive to non-LTE effects \citep{korotin15}, its typical equivalent width is such that, in classical terminology, the line is near or on the the flat part of the curve of growth \citep{dorazi12,redlam15}. The selected Ba\,{\sc ii} 5853 \AA\ line and 4554 \AA\ and 6496 \AA\ lines are the strongest heavy element lines in a star's line list whereas the line selection for La$-$Sm in dwarfs and giants includes several weak lines with EWs $<$\,30 m\AA. 

The outstanding correlation of [Ba {\scs II}/Fe] with stellar activity (Figure \ref{rhk_ba}) with much weaker trends for other $s$-process elements and the greater sensitivity of the Ba\,{\sc ii} line to microturbulence implies the strong Ba {\scs II} lines must have formed in upper photospheric layers whose atmospheric activity is responsible for the changes in chromospheric structure. 

\cite{gurtovenko15} calculated the average depth of formation of the lines under LTE conditions with the empirical solar photospheric model HOLMU \citep{holweger1967,holweger1974}. It is evident from their table that the weak and moderate Fe {\scs I} and Fe {\scs II} lines widely used in the estimation of stellar parameters (T$_{\rm eff}$, log~$g$, $\xi_{\rm t}$) and metallicity of stars have average formation depth very deep in the photosphere with a mean depth of about log$\,(\tau_{5000})=$ $-1.5$\footnote{The optical depth of formation of the line cores log$\,(\tau_{5000})$ corresponds to the optical depth in continuum at $\lambda$ = 5000 \AA\ in the HOLMU model atmosphere.}. It is also clear that all other elements including the $s$-process elements with exception of Ba {\scs II} have an average formation depth close to log$\,(\tau_{5000})$\,=\,$-1.0$ while the strong Ba {\scs II} 5853, 6496 and 4554 \AA\ lines form very far from the continuum forming layers in the upper photospheric layers with average depth log$\,(\tau_{5000})$= $-2.9$, $-4.9$ and $-5.7$, respectively. 

The ID LTE theoretical models such as the {\scs \bf MAFAGS} \citep{fuhrmann93,fuhrmann97} predict an effective line formation depth of log$\,(\tau_{5000})$\,=\,$-$1.9, $-$2.8 and $-$4.7, respectively, for the Ba {\scs II} 5853, 6496 and 4554 \AA\ lines \citep{mashonkina99,mashonkina06}. Such differences in log$\,(\tau_{5000})$ values for the Ba {\scs II} lines is not surprising given that the analyses have employed different model photospheres. The HOLMU model was empirically designed to match the center-to-limb observations of the solar continuum radiation and a variety of spectral lines at different viewing angles ($\mu$)\footnote{$\mu$\,=\,cos\,$\theta$ with $\theta$ describing the angle between the ray direction and the stellar surface normal.}, which confirmed the presence of anisotropic height-dependent photospheric velocity fields in the Sun \citep{holweger1978}. All the 1D model atmospheres (e.g., {\scs \bf ATLAS, MARCS}) based on the assumptions of static atmospheres with no chromospheres and a constant value of microturbulence independent of depth and $\mu$ fail to represent the convective motions in the photosphere realistically \citep{holweger1978,pereira09a,pereira09b}.

The same qualitative behaviour of the 1D models has also been demonstrated for the centre-to-limb behaviour of the Fe lines \citep{pereira09a} and the O I 7777 \AA\ triplet \citep{steffen15}. Although, the three-dimensional (3D) model atmospheres represent both the velocity fields and the thermal structure of the solar atmosphere more realistically, HOLMU model performs much better than all the 1D LTE and non-LTE models \citep{pereira09b,steffen15,lind17}. Therefore, the log$\,(\tau_{5000})$ information provided for the solar Ba {\scs II} lines assuming the depth- and $\mu$-independent microturbulnce velocity in the 1D models ({\scs \bf MAFAGS, ATLAS, MARCS}) may not represent the true depth of formation of Ba {\scs II} lines. These results support the claim, as previously made in \cite{redlam15}, that the greater sensitivity of the Ba\,{\sc ii} line to microturbulence is likely a contributor to resolving the Ba puzzle.

\begin{figure}
\begin{center}
\includegraphics[trim=0.1cm 9.3cm 6.7cm 4.2cm, clip=true,height=0.28\textheight,width=0.73\textwidth]{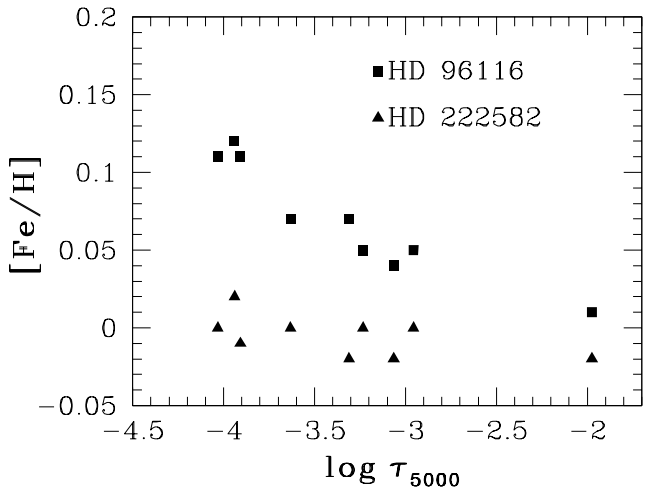}
\caption[]{The differential iron abundance derived from the adopted stellar parameters in Table \ref{abu_solartwins} as a function of the average depth of line formation. }
\label{micro_fe}
\end{center}
\end{figure}

\begin{table}
\caption{Iron abundances for a set of 9 saturated iron lines along with the average depth of formation in stellar photosphere.} 
\vspace{0.2cm}
\centering
\label{depth_feh}
\begin{tabular}{ccccc}   \hline
\multicolumn{1}{c}{$\lambda$} &\multicolumn{1}{c}{$\langle$log$\,(\tau_{5000})\rangle$ } &\multicolumn{2}{c}{[Fe/H] (dex) } \\
\multicolumn{1}{c}{ (\AA) } &\multicolumn{1}{c}{ } &\multicolumn{1}{c}{HD 96116} &\multicolumn{1}{c}{ HD 222582 } \\ 
\hline
 
   4114.4 & -3.065 & +0.04 & -0.02  \\
   4114.9 & -1.974 & +0.01 & -0.02  \\
   5127.3 & -3.632 & +0.07 & \,0.00  \\
   5198.7 & -3.234 & +0.05 & \,0.00  \\
   5225.5 & -2.956 & +0.05 & \,0.00  \\
   5250.6 & -3.310 & +0.07 & -0.02  \\
   5501.4 & -4.032 & +0.11 & \,0.00  \\
   5569.6 & -3.940 & +0.12 & +0.02  \\
   6136.6 & -3.907 & +0.11 & -0.01  \\
   
\hline
\end{tabular} 
\end{table}

To examine this possibility, we have selected a set of 9 iron lines from the Table in \cite{gurtovenko15} whose formation depth is far from the continuum forming layers in the range log$\,(\tau_{5000})=$\, $-2.0$ to $-4.0$. We show in Figure \ref{micro_fe} the differential iron abundances of these lines (Table \ref{depth_feh}) covering the EWs range 60 to 160 m\AA\ in the spectra of HD 96116 and HD 222582 as a function average depth of line formation. All these abundances are measured using the stellar parameters listed in Table \ref{abu_solartwins} whose values are based on the analysis of weak and moderate strength iron lines of EW $<$ 70 m\AA\ \citep{nissen15}. It is evident from Figure \ref{micro_fe} that HD 96116 with relatively high stellar activity has a systematically higher iron abundances for the strong lines formed in the upper layers of the photosphere (log$\,(\tau_{5000})<$\, $-2.0$) than the mean abundance of $+0.01$ dex derived for lines formed deep in the photosphere. Such a trend of iron abundance with depth of line formation strongly suggest a height-dependent microturbulence with velocity increasing toward the top of the photosphere and the transition to the chromosphere. From the very similar iron abundances for the strong Fe lines in HD 222582 suggest that the photospheric structure of the inactive star HD 222582 of age 7 Gyr is very similar to that of the Sun. 

These results strongly suggest that the microturbulence increases steeply with height in the photospheres of active stars relative to dwarfs of age less than the Sun. As a result the application of the microturbulence value derived from the weak and moderate strength iron lines to strong lines such as the Ba {\scs II} lines result in a serious overestimation of the Ba {\scs II} abundances in stars younger than the Sun. For the star HD 96116, an increase of microturbulence by about 0.3 km s$^{-1}$ will drop [Ba/Fe] from the Ba {\scs II} 5853 \AA\ line close to the abundance measured for the other heavy elements La$-$Sm. 
 
\begin{table*}
\begin{center}
 {\fontsize{9}{9}\selectfont
\caption{The mean R$^{\prime}_{5876}$ indices and [X/Fe] ratios for an element X$=$La, Ce, Nd and Sm in local associations of solar metallicity analysed previously in \cite{redlam15}. The number of stars used in calculating the stellar activity index are shown in parentheses.  } 
\label{abu_heavy_associations}
\begin{tabular}{cccccccc}   \hline
\multicolumn{1}{l}{association} & \multicolumn{1}{c}{[Ba {\scs II}/Fe]} & \multicolumn{1}{c}{[La/Fe]} &\multicolumn{1}{c}{[Ce/Fe]} &\multicolumn{1}{c}{[Nd/Fe]} & \multicolumn{1}{c}{[Sm/Fe]} &\multicolumn{1}{c}{R$^{\prime}_{5876}$ (or R$^{\prime}_{HK}$)} & \multicolumn{1}{c}{EW$_{5876}^{(He\,{\scs I}\,D_{3})}$}  \\
\multicolumn{1}{l}{ } &\multicolumn{1}{c}{(dex)} &\multicolumn{1}{c}{(dex)} &\multicolumn{1}{c}{(dex)} &\multicolumn{1}{c}{(dex)} & \multicolumn{1}{c}{(dex)} & \multicolumn{1}{l}{ } & \multicolumn{1}{c}{(m\AA)}  \\ \hline
 
         Argus &$+0.32\pm0.05$ &$+0.04\pm0.04$ &$-0.01\pm0.04$ &   $\ldots$    &$+0.02\pm0.04$ & $-$4.41$\pm$0.03 & 27.43$\pm$2.47(2) \\
   Carina-Near &$+0.32\pm0.05$ &$\,0.00\pm0.05$&$-0.01\pm0.04$ &$+0.10\pm0.05$ &$\,0.00\pm0.03$& $-$4.38$\pm$0.01 & 29.10$\pm$0.30(2) \\
 Hercules-Lyra &$+0.15\pm0.04$ &$+0.09\pm0.03$ &$\,0.00\pm0.03$&$+0.01\pm0.04$ &$+0.01\pm0.02$ & $-$4.56$\pm$0.01 & 17.41$\pm$0.79(3) \\
         Orion &$+0.07\pm0.05$ &$+0.06\pm0.05$ &$-0.05\pm0.05$ &$-0.06\pm0.05$ &$+0.11\pm0.02$ & $-$4.89$\pm$0.07 & 05.25$\pm$1.85(2) \\
   Subgroup B4 &$+0.23\pm0.05$ &$-0.01\pm0.04$ &$+0.05\pm0.04$ &$+0.08\pm0.05$ &$+0.11\pm0.03$ & $-$4.43$\pm$0.03 & 25.85$\pm$2.55(2)  \\
 
\hline
\end{tabular}
  }
\end{center}
\end{table*}

It is well understood from the analysis of the Sun that the chromospherically active phase of the Sun is associated with active upper photospheric layers \citep{schrijver89} and, thus, chromospherically active young stars in the field, OCs and associations have active photospheric layers from where the Ba {\scs II} lines emerge. As a result all those active stars, as found here and in previous studies, have Ba {\scs II} abundances overestimated by the standard LTE abundance analyses where the microturbulence is assumed depth-independent and derived from the weak/moderate strength Fe lines whose formation depth in the photosphere is very deep and differ greatly from the Ba {\scs II} line forming depths. Accordingly, we argue that the Ba abundances determined from the strong Ba {\scs II} lines provide spurious overabundance by the standard LTE abundance analysis of active stars younger than the Sun. Abundances of other elements La$-$Sm represented by weaker lines may be affected, but to a weaker degree (Figure \ref{micro_fe}).

\subsection{Comparison with Clusters and Associations}

The typical strength of the selected Ba\,{\sc ii} 5853 \AA\ line is such that it is readily accessible in OC from FGK dwarfs but the great majority of clusters in the literature, including ours in \cite{reddy12,reddy13,reddy15,reddy16}, have Ba abundances measured from GK red giants. Many giants were chosen over dwarfs because of relatively low $v$\,sin$i$ and high brightness to yield high S/N ratio spectra favourable for accurate abundance determination: sharp lines with strengths from weak to strong for elements sampling the major processes of stellar nucleosynthesis. Furthermore, few of the clusters are local and many were distant with a non-solar metallicity and, thus, extracting the appropriate OC sample with measured Ba abundances and log\,R$^{\prime}_{\rm HK}$ values for comparison with solar twins with [Fe/H] $\simeq 0.0\pm0.1$ dex is a non-trivial exercise. 

The log\,R$^{\prime}_{\rm HK}$ indices of dwarf stars in a large sample of OCs and stellar associations are provided in \cite{mamajek08}, of which three mean-solar metallicity clusters Hyades (0.7 Gyr), NGC 752 (2.0 Gyr) and NGC 2682 (4.3 Gyr) have mean [Ba {\scs II}/Fe] values of +0.30$\pm$0.05 dex, +0.19$\pm$0.08 dex and +0.04$\pm$0.05 dex, respectively, measured from cluster dwarfs by \cite{dorazi09}. The mean log\,R$^{\prime}_{\rm HK}$ values inferred from \cite{mamajek08} for the cluster dwarfs spanning the temperature range of the stars analysed for Ba in D'Orazi et al. are $-$4.50$\pm$0.09, $-$4.7$\pm$0.10 and $-$4.85$\pm$0.11, respectively for Hyades, NGC 752 and NGC 2682. These three OCs of mean-solar metallicity exhibit positive correlation between the values of [Ba/Fe] and log\,R$^{\prime}_{\rm HK}$. 

There exists literature on seven stellar associations but we include in Figure \ref{rhk_ba} only our sample of five nearby young stellar associations -- Argus, Carina-Near, Hercules-Lyra, Orion and Subgroup B4 -- to provide a comparison free of systematic offsets \cite{redlam15} have performed a comprehensive abundance analysis of five local associations whose chemical composition within the measurement uncertainties is very nearly solar for all elements with the exception of Ba which, as we noted in the Introduction, is overabundant by about [Ba {\scs II}/Fe]$=$ 0.2$-$0.3 dex. These associations span the [Fe/H] range $-$0.1 to 0.0 dex comparable to that of solar twins but are very young (ages $\sim$ 3$-$200 Myr). The abundances for all associations are measured from the spectra of FGK dwarfs spanning the T$_{\rm eff}$ range 5150$-$6050 K with stars in the Carina-Near having T$_{\rm eff}$ of 5800 K each.

\begin{figure}
\begin{center}
\includegraphics[trim=0.1cm 10.8cm 8.5cm 4.2cm, clip=true,height=0.22\textheight,width=0.71\textwidth]{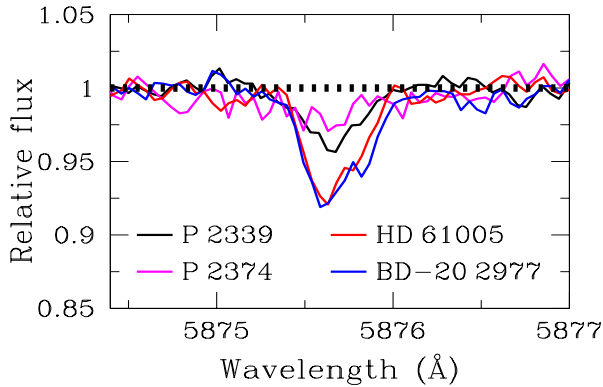}
\caption[]{ Comparison of the He {\scs I} 5875.6 \AA\ D$_{3}$ profiles in the spectra of stars in Orion (Parenago 2339 \& 2374) with stars in Argus (HD 61005 \& BD-20\,2977). The weaker He {\scs I} profiles in Orion stars suggest that their chromospheric activity levels are much less than the value obtained for the Argus association. The dashed line represents the continuum level. }
\label{hed3_asso}
\end{center}
\end{figure}

We removed the telluric absorption by dividing the stellar spectra with a rapidly rotating hot star spectra and the EWs of He {\scs I} 5876 \AA\ D$_{3}$ lines were measured. All the spectra correspond to S/N ratios of 200$-$350 around the He {\scs I} D$_{3}$ line whose EWs were transformed to R$^{\prime}_{\rm HK}$ values using equation \ref{eq_ewvsrhk}. The computed mean values of R$^{\prime}_{\rm HK}$ indices along with the average Ba abundances of FGK dwarfs in associations are listed in Table \ref{abu_heavy_associations}. 

A comparison of the He {\scs I} D$_{3}$ profiles in the spectra of two stars each from the Orion and Argus associations is displayed in Figure \ref{hed3_asso}. Both these associations are very young with ages of 3 Myr \citep{walker1969,blaauw1954} and 30 Myr \citep{desilva13} for Orion and Argus, respectively. The lithium abundances derived previously for the stars in Orion \citep{cunha1995,redlam15} and Argus \citep{redlam15} also confirm the youth of these associations. But, as seen in the Figure \ref{hed3_asso}, the He {\scs I} D$_{3}$ lines in the stars Parenago 2339 and Parenago 2374 of Orion are much weaker than the corresponding profiles in members (HD 61005 and BD-20\,2977) of the Argus association. The large scatter of about 0.5 dex in log\,R$^{\prime}_{\rm HK}$ found for stars at the young ages of a few Myr suggests that young stars may have larger intrinsic variations in activity levels.

We have checked whether the log\,R$^{\prime}_{\rm HK}$ of $-$4.41 derived from the two stars (HD 61005 and BD-20\,2977) in Argus associations agrees with such estimates in the literature. To our knowledge, only the star HD 61005 has a measured value of log\,R$^{\prime}_{\rm HK}$\,=$-$4.31 from the Ca {\scs II} H and K emission line fluxes \citep{desidera11}. The satisfactory agreement of measured log\,R$^{\prime}_{\rm HK}$ values between the analyses not only supports the reliability of our log\,R$^{\prime}_{\rm HK}$ values obtained from the He {\scs I} D$_{3}$ lines but also confirms that the stars in Orion have intrinsically lower value of log\,R$^{\prime}_{\rm HK}$ for their age. Therefore, the log\,R$^{\prime}_{\rm HK}$ values for stars in these associations supports, as demonstrated previously for solar twins in \cite{ramirez14}, the presence of larger intrinsic variations in stellar activity levels among the young stars.

\begin{figure}
\begin{center}
\includegraphics[trim=0.0cm 2.7cm 8.5cm 4.2cm, clip=true,height=0.4\textheight,width=0.63\textwidth]{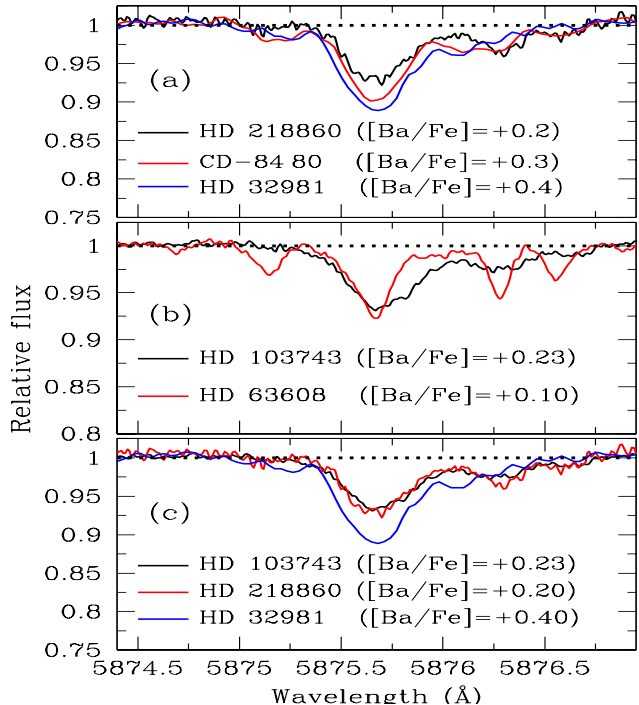}
\caption[]{ Comparison of the He {\scs I} 5875.6 \AA\ D$_{3}$ profiles among the members of AB Doradus (Panel a), among the members of Carina-Near (Panel b) and between stars of AB Doradus and Carina-Near associations (Panel c). The chemical content of [Ba\,{\scs II}/Fe] in stars is also displayed in each panel.} 
\label{hed31_dorazi}
\end{center}
\end{figure}

We probe through the Figure \ref{hed31_dorazi}, the variation of [Ba/Fe] from the Ba {\scs II} 5853 \AA\ line with the strength of He {\scs I} D$_{3}$ profiles (level of stellar activity) in stars of the associations AB Doradus (70 Myr) and Carina-Near (200 Myr) analysed previously by \cite{dorazi12}. Our interest in such a comparison (Figure \ref{hed31_dorazi}) is occasioned by the reported internal chemical inhomogeneity of [Ba\,{\scs II}/Fe] values relative to uniform La and Ce abundances among members of a given association (Table 1 in \citealt{dorazi12}). The readily available multiple high-resolution FEROS (F) and HARPS (H) spectra of the stars HD 32981 (F), CD-84\,80 (F) and HD 218860 (H) in AB Doradus, and the srars HD 63608 (H) and HD 103743 (H) in Carina-Near association were retrieved from the ESO public archive\footnote{\url{http://archive.eso.org/eso/eso_archive_main.html}}. The S/N ratio of the co-added spectra at 5880 \AA\ runs from about 300 (HD 32981, HD 218860) to 750 (CD-84 80, HD 63608, HD 103743). All these stars were explored previously for the abundances of 11 elements in the range Li to Zn by \cite{biazzo12}, and for the heavy elements Y, Zr, Ba, La, and Ce by \cite{dorazi12} whose [X/Fe] values for any element X exhibit solar ratios with the sole exception of Ba where the [Ba/Fe] values are overabundant by 0.2 dex with an intrinsic abundance spread of about 0.2$-$0.3 dex in [Ba\,{\scs II}/Fe] among members of a given association (see Table 1 in \citealt{dorazi12}). 

Figure \ref{hed31_dorazi} displays the comparison of He {\scs I} D$_{3}$ line strengths among stars of AB Doradus (panel a), among members of Carina-Near (panel b) and between the members of AB Doradus and Carina-Near (panel c). In each panel, we provide the [Ba\,{\scs II}/Fe] content of the stars measured previously by \cite{dorazi12} for the Ba {\scs II} 5853 \AA\ line using the high-resolution and high S/N ratio stellar spectra. Panel (a) displayed for the members of AB Doradus confirms the existence of a correlation between the line strength of He D3 profile (and, hence, the level of stellar activity) and [Ba/Fe] values to the level that is discernible among stars of a given association i.e., among stars belong to the same association, the star with relatively high level of chromospheric activity is associated with relatively high [Ba/Fe] values. Inspection of panel (b) in the Figure \ref{hed31_dorazi} confirms that the trend of Ba II abundances with chromospheric activity seen among members of AB Doradus is repeated for the members of Carina-Near association. Finally, we compare in panel (c) the He {\scs I} D$_{3}$ profiles in stars of AB Doradus (HD 32981, HD 218860) with the Ba-enriched star HD 103743 of Carina-Near association. The primary evidence that comes from the panel (c) in Figure \ref{hed31_dorazi} is the association of very similar measured [Ba/Fe] values for stars belong different associations to identical strengths of the He {\scs I} D$_{3}$ profiles. The repetition of the correlation of [Ba\,{\scs II}/Fe] with chromospheric activity (line strength of the He {\scs I} D$_{3}$ profiles) among coeval group of stars in a given stellar association reinforce that the strengthening of the Ba\,{\scs II}/Fe] line and, thus, the [Ba\,{\scs II}/Fe] overabundance seen at young ages ($<$ 100 Myr) is associated with the stellar activity but not with the stellar age. This novel result from our study evidently reject previous reports of enhanced $s$-process nucleosynthesis of Ba with decreasing Galactic age \citep{dorazi09, dorazi17,mishenina15} and, thus, the Ba overabundance measured for young stars, OCs and associations is unrelated to aspects of stellar nucleosynthesis and Galactic chemical evolution.

A graphical comparison of the mean [Ba/Fe] with the mean stellar activity index of the associations is provided in Figure \ref{rhk_ba}. The typical star-to-star abundance scatter within the association is very small such that the mean [Ba/Fe] values have a dispersion of about 0.05 dex, a value slightly larger the scatter in [Ba/Fe] found for the solar twins. 

Inspection of the [Ba/Fe] versus log\,R$^{\prime}_{\rm HK}$ for the stellar associations confirms that the trend of Ba {\scs II} abundances with chromospheric activity seen among the solar twins is repeated for the local stellar associations. The significance of this distinct trend for Ba {\scs II} is boosted by the independence of log\,R$^{\prime}_{\rm HK}$ values with the age of associations (Figure \ref{rhk_age}). More importantly, the youngest stellar association Orion with an age of 3 Myr has a measured [Ba/Fe] of $+$0.07$\pm$0.05 dex, a value appreciably less than $+0.6$ reported in \cite{dorazi09} for clusters of age $<\,$50 Myr. These associations further demonstrate that the spread about the [Ba/Fe] trend is compatible with identical spread seen for the solar twins and, thus, integrate well into the indisputable trend of increasing [Ba {\scs II}/Fe] with increasing chromospheric activity. The range $-$0.06 to $+$0.1 dex in abundances covered by La, Ce, Nd and Sm relative to Fe in associations is comparable to [X/Fe] range of La$-$Sm in solar twins and, thus, the La$-$Sm abundances in associations also confirm that they are independent of stellar activity. The repetition of the positive correlation of [Ba {\scs II}/Fe] with chromospheric activity indices for the young associations and OCs reinforces the impression that exceptionally high [Ba {\scs II}/(La$-$Sm)], as we noted in the Introduction, of up to $+$0.3 dex in the young F-G dwarfs of local associations and up to $+$0.4 dex in OCs is unrelated to aspects of Galactic chemical evolution but result from the overestimation of Ba {\scs II} abundances by the standard LTE abundance analysis of stars in the youngest OCs and associations. The trend in Figure \ref{rhk_ba}, as explained in previous section, likely signals the influence of turbulent upper photospheric layers in strengthen the Ba {\scs II} 5853 \AA\ lines in active stars where the upper photospheric layers are characterized by stellar activity. Such a stellar turbulence is lessened in inactive and older stars as the level of stellar activity decays with time. 

In a recent study using the high-resolution ($R\,=$\,48,000) spectra of pre-main sequence dwarfs in young clusters ($<$50 Myr), \cite{dorazi17} found that the heavy elements La and Ce exhibit solar ratios with [Ba/Fe] from the Ba {\scs II} 5853 \AA\ line being overabundant by up to 0.65 dex. They have discussed several possible explanations such as the chromospheric effects, departures from the LTE approximation and the $i$-process nucleosynthesis and suggested that the last explanation is likely the source of Ba overabundance. But they ``cannot currently provide the definite answer'' which we describe in this paper via an extensive analysis of the solar twins and stellar associations. Our novel results in this study do not support previous reports of enhanced production of Ba via the enhanced $s$-process nucleosynthesis or by the operation of neutron-capture $i$-process in AGB stars with decreasing Galactic age \citep{dorazi09, dorazi17,mishenina15} and, thus, the Ba overabundance measured for young stars, OCs and associations is unrelated to aspects of stellar nucleosynthesis and Galactic chemical evolution.

Barium abundance determinations  have been extended to pre-main sequence stars in the Lupus T association with an age of about 4 My with \cite{biazzo17} reporting [Ba/Fe] $\sim +0.7$ for the warmest or K-type stars with T$_{\rm eff}\sim\,5000$ and $\log g \sim 4.0$. This result appears to extend a trend of increasing [Ba/Fe] with decreasing  stellar age among associations and open clusters. The abundances of Fe and Ba were obtained from spectra at a resolution of only $R \sim 8,800$ and at this resolution abundances could not be obtained for other heavy elements, all represented by weaker lines than their Ba\,{\sc ii} counterparts. At present, it is unclear if the Ba abundances of the Lupus pre-main sequence stars are truly reflecting the composition of the stellar atmospheres or are enhanced by measurement errors or  have been systematically overestimated through a failure of classical model atmospheres to represent the real atmospheres, specifically as suggested in this paper, by a mischaracterization of the microturbulence around the photosphere-chromosphere boundary. These Lupus stars are decidedly the example {\it par excellence} where improved spectra are now required! Then, it should be possible to understand better the Ba puzzle provided by the very youngest of stars.

\section{Concluding Remarks}
In this paper, we have presented elemental abundances for the suite of heavy elements Ba, La, Ce, Nd and Sm with La$-$Sm are analysed for the first time in a sample of 24 solar twins. With the main goal of investigating the puzzling overabundance of Ba at young ages in solar twins, OCs and local associations, we have exploited high-resolution, high S/N ratio spectra of the solar twins and measured the heavy elements and chromospheric activity indices on a homogeneous basis, consistent with our previous studies. 

The principal novel results presented in this paper are (i) the Ba {\scs II} line abundance correlates positively with the value of stellar activity index (log\,R$^{\prime}_{\rm HK}$, Figure \ref{rhk_ba}) while other heavy elements from La to Sm are consistent with the claim that [X/Fe] is unchanged with stellar activity. Similar trends among the heavy elements are seen among stellar associations and OCs with [Fe/H] $\sim 0.0$, as anticipated. (ii) the indisputable trend of increasing [Ba\,{\scs II}/Fe] with stellar activity seen for the solar twins is also discernible among dwarf stars within stellar associations (Figure \ref{hed31_dorazi}) whose members are expected to be coeval.

The overabundance of Ba {\scs II} relative to La$-$Sm in OCs and associations has been reported previously by us and others but we are the first to demonstrate that the barium overabundance provided by Ba\,{\sc ii} lines in young stars, associations and OCs is not nucleosynthetic  in origin but strongly associated with the level of stellar activity. Unlike the strong Ba {\scs II} line forming depths (log$\,(\tau_{5000})$= $-2.9$ to $-5.7$), the average depth of formation for all elements including the weak lines of heavy elements Ba {\scs I}, La$-$Sm is very deep in the photosphere and close to the continuum forming layers. Within the classical picture of LTE abundance analysis, a constant value of microturbulence velocity derived from the weak/moderate strength Fe lines that have formed deep in the photosphere (log$\,(\tau_{5000})\lesssim$ $-1.5$.) is insufficient to represent the true broadening imposed by relatively turbulent upper photospheric layers on the Ba {\scs II} lines. Thus, the inclusion of microturbulence derived from Fe lines in the spectrum synthesis of strong Ba {\scs II} lines results in a serious overestimation of Ba abundance. The strong correlation of [Ba {\scs II}/Fe] with the chromospheric activity is an indication that the increased stellar turbulence in the upper photospheric layers of active stars provide significant overestimated Ba {\scs II} line abundances relative to stars with log\,R$^{\prime}_{\rm HK}$ close to the Sun.

We demonstrate clearly that the puzzling overabundance of [Ba {\scs II}/Fe] in young dwarfs, associations and OCs is unrelated to aspects of $s$-process nucleosynthesis in AGB stars and Galactic enrichment via a long-forgotten neutron-capture $i$-process \citep{mishenina15} but has returned from the overlooked extra broadening of Ba {\scs II} lines in young stars where the Ba {\scs II} line forming layers are more turbulent than the microturbulence value derived from the LTE analysis of weak/moderate strength Fe lines buried deep in the stellar photospheres.

\vskip1ex 
{\bf Acknowledgements:}

We thank Valentina D'Orazi for a comprehensive and thoroughly helpful referee's report and Poul Nissen for his most helpful comments on a draft of this paper. DLL wishes to thank the Robert A. Welch Foundation of Houston, Texas for support through grant F-634. 
Based on data products from observations made with ESO Telescopes at the La Silla Paranal Observatory under programs 072.C-0488, 074.C-0364, 088.C-0323, 183.C-0972, 188.C-0265, 192.C-0224.


\end{document}